# A solution for reducing high bias in estimates of stored carbon in tropical forests (aboveground biomass)

**H. Arellano-P.[1,2] and J. O. Rangel-Ch.[1]**

[1]{Grupo de Investigación en Biodiversidad y Conservación, Instituto de Ciencias Naturales, Universidad Nacional de Colombia, P.O. Box 7495, Bogotá D. C., Colombia}

[2]{Compensation International Progress S.A., P.O. Box 260161, Bogotá D. C., Colombia}

Correspondence to: H. Arellano-P (harellano@unal.edu.co)

## Abstract

**A nondestructive method for estimating the amount of carbon stored by individuals, communities, vegetation types, and coverages, as well as their volume and aboveground biomass, is presented. This methodology is based on information on carbon stocks obtained through three-dimensional analysis of tree architecture and artificial neural networks. This technique accurately incorporates the diversity of plant forms measured in plots, transects, and relevés. Stored carbon in any vegetation type is usually calculated as half the biomass of sampled individuals, estimated with allometric formulas. The most complete of these formulas incorporate diameter, height, and specific gravity of wood but do not consider the variation in carbon stored in different organs or different species, nor do they include information on the wide array of architectures present in different plant communities. To develop these allometric models, many individuals of different species must be sacrificed to identify and validate samples and to minimize error. It is common to find cutting-edge studies**

**that encourage logging to improve estimates of carbon. In our approach we replace this destructive methodology with a new technique for quantifying global aboveground carbon. We demonstrate that carbon content in forest aboveground biomass in the pantropics could rise to 723.97 Pg C. This study shows that a reevaluation of climatic and ecological models is needed to move toward a better understanding of the adverse effects of climate change, deforestation, and degradation of tropical vegetation.**

## Introduction

Significant uncertainty in the calculation and estimation of biomass and carbon stored in tropical forests (Baccini *et al.* 2012, Chave *et al. 2014*) is an issue of great concern to the scientific community. Usually, governmental and nongovernmental entities responsible for developing and implementing policies and mechanisms to mitigate climate change, such as REDD+ (Reducing Emissions from Deforestation and forest Degradation), are also stakeholders. To estimate the losses or gains of multitemporal carbon for a given area, two key inputs are necessary: thematic mapping coverage for at least two sampling periods, and calibrated biomass and carbon estimates obtained using various methods. The most common method used worldwide for estimating forest biomass is the sum of biomass values for sampled individuals, determined with an allometric model. To develop this type of model, it is necessary to harvest many individuals of various species (Chave *et al. 2014,* Feldpausch *et al. 2012,* Chave *et al. 2005*) to determine and validate species and to minimize error. Weight information for 2,410–4,004 tropical trees, cut down and segmented, was required for the allometric equations used in just these three cited studies (Chave *et al. 2014*, Chave *et al. 2005,* Overman *et al.* 1994). However, biomass results obtained with traditional methods have an unacknowledged bias (Chave *et al. 2005*) toward individuals with a diameter at breast height (DBH) of less than 5 cm or greater than 156 cm (Chave *et al. 2005*) to 212 cm (Chave *et al. 2014*). As a consequence, the results often lead to misinterpretations of the statistical coefficient of determination ($r^2$) (Sileshi 2014), creating a false impression of the reliability of the models. This situation also occurs in other proposed equations (Feldpausch *et al. 2012,* Overman *et al.* 1994, Sileshi 2014), mainly because of the mathematical structures they share. In many cases, only the constants were different among the equations. We believe that the traditional methodology (Baccini *et al.* 2012, Chave *et al. 2014,* Feldpausch *et al. 2012*) is unreliable, has a high economic cost, increases unwanted deforestation, and leads to errors in the equations resulting from cutting and weighing methods. In addition, building a database large enough to increase accuracy would take more than 50 years (Chave *et al.* 2005). The importance of carbon held in biomass is recognized by the scientific community (Baccini *et al. 2012,* Chave *et al. 2014,* Feldpausch *et al. 2012,* Asner 2009). However, although the amount of carbon held ranges between 43% and 58% of the total biomass (Schulze *et al.* 2000), because of floristic composition, tissues and organs examined, stand age, population density of each stratum, and other variables within plant communities, the Intergovernmental Panel on Climate Change (IPCC) promotes the use of a 50% figure in calculating carbon reserves (IPCC 2006).

To address this concern, a new, nondestructive methodology using three-dimensional modeling based on the species present in typical plots, their organs, and species- and organ-specific carbon storage data derived from small tissue samples was developed as an alternative to improve the calculation of biomass and carbon. Field information on different vegetation types established in a mesoclimate tropical gradient in the Colombian Caribbean was collected over four years. This dataset, together with knowledge of these species' regional distributions, enabled us to replace allometric equations with a new method based on three-dimensional modeling to estimate volume using various organs of multiple individuals from different plots. This new technique estimates the amount of carbon held in the tropical forest aboveground biomass with overwhelming accuracy (error values were between 1.79% and 4.03%), and replaces the destructive techniques promoted in the past (Chave *et al. 2014*, Feldpausch *et al. 2012*). We were able to calibrate the pantropical carbon estimates to date, allowing us to compare our method with two previous estimates (Baccini *et al*. 2012, Feldpausch *et al.* 2012) based on the ratio of forest aboveground biomass to total pantropical carbon. Current models to estimate biomass and carbon must be revised and improved by monitoring forest dynamics. Core variables (e.g., climate, soil, nutrient content, topography, disturbance history, and socioeconomic factors) should be included and combined with spatially specific algorithms to explain regional and local differences. An accurate dataset for these models also requires a large-scale remote-sensing system, and all these outcomes must be connected with a strong network of fieldwork and lab techniques. Only with the combination of these procedures will we be able to evaluate the degradation, and perhaps recovery, of the vegetation and its connections with biodiversity and the carbon cycle. If monitoring strategies improve, the role of forest dynamics in climate-change mitigation and adaptation and in the carbon cycle will be better understood. Less uncertainty in carbon-cycle models, including their impacts on biodiversity, will be essential to the success of forest governance policies, in particular the REDD+ initiatives (Bustamante *et al*. 2016).

## Study area

In an area of 3,425,083 ha located in the southern and northwestern regions of the department of Cordoba, in the Colombian Caribbean (datum UTM-WGS 84 zone 18 N 78-72 W) 1,030,572 North, 835,192 South, 348,788 West and 529,688 East), the structures of 27 forest types along an altitudinal gradient from 0 to 1,839 m were analyzed. To collect the field information, 112 plots of 500 m² were set up, 6,141 individuals with DBH of ≥ 1 cm were measured, and 938 species were identified. With this information, floristic compositions of 27 forests and 27 types of tall shrubby communities were classified, and 5 combinations of coverages where agricultural use predominated were differentiated. For the carbon sampling, 4 forests were chosen. These were dominated and characterized by *Jacaranda copaia* and *Pouteria multiflora; Protium aracouchini* and *Virola elongata; Trichilia hirta* and *Schizolobium parahyba;* and *Acalypha diversifolia* and *Guazuma ulmifolia*, respectively (Avella & Rangel 2012, Rangel *et al.* 2010). These communities occurred along a tropical climate gradient from a superhumid climate with rainfall between 2,200 and 3,000 mm per year to a semihumid climate with rainfall between 1,000 and 1,400 mm. It is important to note that these vegetation types contain the information necessary for inferring characteristics of other vegetation assemblages.

## Methodology

An examination of traditional methodologies used to quantify biomass and carbon in tropical regions (Chave *et al.* 2014, Feldpausch *et al.* 2012, Chave *et al.* 2005, Overman *et al*. 1994), applied for over half a century (Ketterings *et al.* 2001), revealed that several practices could be replaced with the help of technological tools. Traditionally, to calculate the weight of a tree it had to be cut down; however, this is no longer necessary with the use of techniques to calculate the precise volume by three-dimensional modeling of the tree’s organs (relating the values obtained to their respective densities) and to estimate their weight. Thus, individual plants sampled in the studied area were modeled with the combined use of drawings, photographs, Bezier curves, NURBS (Non-Uniform Rational B-Spline), and the latest surface algorithms.

Routines available in commercial software packages (SpeedTree studio 6.1.1 and Xfrog 4.5 for Maya 2009) were used because they contain advanced organic modeling and dimensioning tools. They also have the capacity to generate different levels of surface quality and recreate trees very quickly. The process consists of combining basic structural information (DBH, height, and coverage) with data from statistical counts, field illustrations, photographs, level of branching, number of branches, branch thickness, and adjusted distance metrics to generate a matrix model XYZ in normed vector space $\boldsymbol{R}^3$. To link carbon records for each of the sampled plant organs, the final branching three-dimensional model was separated into two groups, one that contained thick stems and another that contained young branches and their bark. The volume was calculated by numerical integration (Mirtich 1996, Gosline 1997) and voxelization (Patil & Ravi 2005, Aitkenhead 2011) in 221 individuals (belonging to 127 species) represented by three-dimensional models. The DBH of modeled individuals ranged from 1.83 to 195 cm. The corrected height of the evaluated models was 45.19 m, with lower and upper limits of 2.01 m and 47.92 m. Various software packages with three-dimensional metric manipulation capabilities (Maya, Surface, and Rhino3d, among others) were used for measurement comparisons and showed the correct bounding of the models. According to a preliminary DBH classification, 174 individuals were recorded in the range of 1.83 to 50 cm. Of these, more than half were under 30 cm. Between 50 and 100 cm, 32 individuals were modeled, of which more than half do not exceed 70 cm in diameter. Between 100 and 150 cm, and between 150 and 197 cm, 5 and 4 individuals were modeled, respectively.

The biomass of each individual was obtained by adding the wet and dry density values sampled in the stems from the main and secondary branches, the bark, and the leaves. Next we compared biomass with carbon content of the samples within and among species, taking into account that carbon concentration in the organs is variable (Elias & Potvin 2003, Lamlom & Savidge 2003, Zhang *et al.* 2009). A carbon concentration between 41.11% and 49.99% of total biomass was recorded (this analysis was conducted in part by Lozano 2012). To incorporate morphological heterogeneity of the vegetation, after recording total carbon for individuals we classified them by their tree architecture (Keller 2004). Calibration models were built in the form of an artificial neural network, with training matrices for different intervals of DBH, combined with type of architecture, DBH records, height, and specific gravity of the wood. The response variable was amount of carbon stored in the biomass, calculated using the three-dimensional modeling-based methodology. Using these methods, stored carbon was adequately calibrated in 5,920 individuals with typical structural information. In the final count, information from 6,141 individuals was consolidated. The overall correlation coefficients of the processes ranged from $R = 0.73$ to 0.97, with less-frequent architectures having lower values, as expected. Subsequently, carbon stored in different vegetation types was calculated and compared with values obtained using allometric equations (Chave *et al.* 2005, Overman *et al.* 1990). To calibrate the density of individuals per area unit, and to extend the observations to areas greater than 1 ha, we analyzed floristic similarities using principle coordinates (MDS), Sokal and Michener’s distances, and structures using canonical analysis of populations (MANOVA). These analyses included dominant species of the arboreal strata belonging to different vegetation types established in Colombia, Panama, and Costa Rica (Gentry 1988, Boyle 1996). Next we produced a map of the southwest Colombian Caribbean combining information on vegetation types with regional estimates of total carbon content (Arellano 2012). Finally, to calibrate pantropical carbon, several statistical tests were performed to determine the use of training variables such as elevation, annual amount of precipitation, monthly average temperature, pantropical carbon (Baccini 2012), and tree height (Simard *et al.* 2011). The best results were obtained with the last two variables.

**Biomass calculation using the three-dimensional species model and tissue density**

For a better understanding of the three-dimensional processes performed through the normed vector space $R^3$, see Fig. 1. There are two methods for calculating the volume of three-dimensional objects with complex surfaces: the numerical integration method, developed by Mirtich (1996), and object voxelization, outlined by Patil & Ravi (2005). For the correct calculation of the volume of individuals using numerical integration (vector method), it is necessary obtain a homogeneous and topologically closed polygonal mesh. In contrast, the voxelization method (three-dimensional raster method) requires only a closed mesh, independent from the arrangement of vertices in triangles, squares, or other two-dimensional geometric configurations in the normed vector space $R^3$. The advantage of the latter is that a homogeneous mesh is not necessary.

In general, the numerical integration method consists of three steps, in which the volume vector components are simplified through the application of divergence, projection, and Green theorems (Mirtich 1996). In this contribution, we used the adaptation for MATLAB 2011b (developed by Andrew Gosline) of the C code programmed by Mirtich (1996), and we modified part of the code to increase computation speed and start the batch file process. The method also makes it possible to correct records of individual heights (in several cases by keeping the diameter at breast height [DBH] record intact and comparing it with the scaled records graphics); to obtain information about growth angles, leaf-insertion angles, branch density, and density of projected shadows at different times of the day; and to calculate the mass center, mass momentum, inertia tensor, surface, actual coverage, volume, and effect of natural and anthropogenic disturbances affecting tree growth.

Voxelization creates a continuous geometric representation using a set of voxels that reproduce a real object with maximal accuracy (Patil & Ravi 2005). Voxels, like their two-dimensional counterparts, pixels, can be manipulated by choosing the desired cubic unit to define the overall resolution of the object. That is, the representation of trees by voxels using current technology could be done at micrometer, millimeter, or centimeter scales. The choice of one or the other mainly depends on constraints imposed by the hardware configuration used. One cubic centimeter was selected as a unit for individuals under 15 m in height; 1.95 $cm^3$ for some individuals between 15 and 20 m, and coverages with sides shorter than 10 m; 8 $cm^3$ for individuals with heights between 15 and 30 m; and 15.62 $cm^3$ for individuals over 30 m. These units were chosen because the voxelization method generates a logic matrix that emulates, in small compartments, the space that a determined object and its complement occupies (empty space), to fill a hexahedron. This means that at the centimeter scale, the representation of structures that fit a cube of 15 m on a side generates about 3,375,000,000 records, a situation in which, by doubling the height, the amount of memory space required (RAM) is cubed, as is the time required to perform the calculation on a computer. Representing all the trees in small units would require an excessive amount of time, as well as the generation of a new computational code to split large files or the implementation of a computer farm for their calculations. In contrast, choosing the largest unit (15.62 $cm^3$) for analyzing all modeled trees would adequately represent large volumes and even compensate calculations of un-modeled structures smaller than a 6.452 $cm^2$ in cross-sectional area. However, for individuals with lower heights, the error would increase in inverse proportion to its height. For this reason, all individuals were voxelized with logical matrices that did not exceed 3,375,000,000 of records, using the units specified above. In addition, correction factors were incorporated, calculated by comparing similar volumes in different cubic units. Voxelization methods include scanning conversion, paired counts, distance projections, and light-ray simulation. In this contribution, ray tracing simulation was used, in which intersections were found between a simulated light beam (fired orthogonally to the modeled object), for each of the X, Y, and Z axes, and the sides of the object, in the normed space $R^3$. The mathematical explanation of the method is given in Appel (1968) and Whitted (1980). It is important to emphasize that the length of each axis corresponds to the actual dimension of the object in the chosen unit of measure. These dimensions are found by determining the difference between the maximums and minimums (boundaries limits) of the vector model coordinates. This calculation involves three rays' simulations, using geometric equations, of the path of a light ray that intersects a vector object (vertices, normals, and faces) from a light source to a point on the plane of the image or screen. To increase computing speed, the process is reversed, and any calculation that does not generate an image record on the plane is deleted. Logical values (true or false) along the path of each simulated ray for each voxel projected on the screen generate a three-dimensional logic matrix. This contains existential information in the $R^3$ space of the object. These records do not contain positions in $X, Y, Z$ space; rather, their positions were deduced by the place occupied by a real value in the matrix and the voxel size (Patil & Ravi 2005). Thus, it is possible to logically represent complex objects and calculate their volumes using the continuous sums of the evaluated spaces. In Figs. 2, the most important characteristics of a homogeneous triangle mesh are shown, which are necessary for making estimates using numerical integration. The three-dimensional model can be used to correct records of individual heights (in several cases, measurements were corrected by comparing the diameter at breast height [DBH] with the proportionality of the tree photographs with digital projection boundaries). This technique produces information about growth angles, leaf-insertion angles, branch density, and density of projected shadows at different times of the day and can be used to calculate the mass center, mass momentum, inertia tensor, surface, actual coverage, volume, and effect of natural and anthropogenic disturbance on tree growth. Figure 3 shows two of the tree models constructed by voxelization. To generate these models, part of the code (Aitkenhead 2011) was modified to increase computational speed, calculate the volume and dimensions of modeled trees, and generate batch files. To identify any inconsistencies, the measurements surveyed by the models were evaluated with different software (Maya, Surface, and Rhino3D).

Palm stems were included in this analysis as deformed cylinders. For species that retained part of the leaf sheaths, estimates for them were included in leaf models. The first level of branches, those coming from the trunk or stem, were collectively considered the main branching. The third, fourth, fifth, and sixth levels represented secondary branching, the highest levels being the youngest and thinnest.

After the detailed estimate and calibration of tree architecture volume without leaves or roots, we proceeded to estimate branch volume above the third level, following the same method. This was done to differentiate volumes of organs, first by thickness and then by age. Equation 1 estimates the volume (a right prism with trapezoidal base) through the relation between the total surface area of the individual (estimated without taking into account existing transverse regions) and branch surface area with their respective bark thicknesses. Some structures present in the bark, such as surface shapes, rings, or spines, were modeled according to the increase that these generated in the computer algorithms. The size of the file depends on complexity of modeled structures, format used, density of the triangle or square mesh, and the generalization processes. The STL (stereo lithography) format was chosen because it stores the topological structure efficiently as a text file (vertices, normals, and faces indexing). Thus small files (about 60 MB on a disk) can store large amounts of information. A file of this size can hold the 2,630,322 vertices, 2,630,322 normals, and 292,258 faces representing an individual of *Cedrela odorata* with a DBH of 19.51 cm and a height of 10.35 m. A more complex individual of *Caryocar amygdaliferum* with a DBH of 123.75 cm and a height of 41.71 m requires about 300 MB (13,239,287 vertices).

**Equation 1**. Bark volume ($V_B$) is estimated for main and secondary branches, based on initial and final bark thickness, $B_i$ and $B_f$, and the surface of organs, $Su$ (volume of a straight prism with trapezoidal base).

$$V_B = \left( \left(B_i + B_f\right)\left(Su^{1/2}\right) / 2 \right) Su^{1/2}$$

With the volume values found by numerical integration and voxelization we proceeded to calibrate the models, especially those generated by voxelization with units other than cubic centimeters. Volumes of stems, main branches, secondary branches, and bark were multiplied by their respective green density records to obtain hydrated biomass values for individuals (Vásquez & Arellano 2012). With the biomass record of each organ evaluated, we next estimated the reserves of structural carbon, nitrogen, and hydrogen.

**Methodology for leaf treatment**

Estimates of volume related to biomass and carbon contributed by leaves were calculated in a different way, which can be summarized in three steps: a statistical count of the number of leaves on different branches, a detailed estimate of surface area, and relation of the latter with leaf thickness. To estimate number of leaves in a particular individual, SpeedTree studio 6.1.1 software was used. This program incorporated minimum and maximum organ counts made at each branch level, using a motor that generates a core of pseudorandom numbers within set limits. A matrix of number of leaves per branching level was obtained, whose values were multiplied by the most frequent volume estimates for leaves at the various levels. To estimate leaf surface area, we used photographs or ink silhouette drawings of leaves collected in the field. With the digital information in a raster format, mosaics were generated, which grouped together about 47 sheets of paper in a 1:1 scale, using Adobe PhotoBridge software. In GIMP software, noise and drawing errors were eliminated, silhouettes were filled with color, and RGB information was converted in a binary model of one bit per pixel (black and white). In this step, dimensions of the mosaic were bounded, identifying the exact measurements of each side. This information was used to reference and position the resulting vector model. Next, data were exported in files with referencing capabilities (such as TIF and TIFF), for geometric correction in any geographic information system (using GRASS 6.4 software). If an export data file does not have referencing capabilities, it is still possible to accomplish this step, incorporating the TIC information of the resulting vector file. With the same software, the binary export file in TIF format is projected again under the Transverse Mercator coordinate system, with WGS84 datum, a scale factor of 1 (one), and parameters with 0 (zero) values. This guarantees the submetric accuracy necessary to adequately represent leaves in vector format. With the GRASS 6.4 command r.to.vect, and the area option enabled, we obtained the vector model of drawn objects for further editing, and the creation of area and perimeter attributes. The standard output of this process is shown in Figure 4.

It is important to note that information from leaves combined or divided to simplify their illustration should be carefully reconstructed for the final matrices. This ensures that the leaf area corresponds correctly to its record and not to only a part or leaflet of it. An additional benefit of this procedure is the ability to analyze, with any geographical information system, the estimated leaf records, employing the same methodologies used in landscape ecology. Measurements such as fractal dimension, perimeter-area ratio, complexity, core analysis, and shape indices, among others, are equally applicable at this level. The outcomes generated a matrix that summarized the two-dimensional leaf morphology. Multiplying element by element, leaf-thickness records were incorporated to obtain the final volume matrix. This matrix was multiplied by the leaf-count matrix to obtain, as an intermediate result, the foliar volume by size group contributed by each individual. The preceding steps were followed for the other plant organs, with the difference that the leaf-volume matrix was multiplied, element by element, by the estimated hydrated and dry density corresponding to each calculated volume. Sums for the rows and finally total estimates of hydrated and dry biomass were obtained for each individual.

Total surface area of the leaves was estimated by multiplying leaf count by the area of each leaf. The total area of each leaf is found by adding the areas of the two sides and the product of leaf perimeter and thickness. This yields a precise estimate of the area involved in gas exchange, creating a breakthrough for future understanding and calibration of atmospheric variables such as water vapor, gas exchange, respiration, and transpiration. Finally, leaves of the palm species *Astrocaryum malybo, Attalea butyracea, Oenocarpus bataua, Sabal mauritiiformis, Socratea exorrhiza,* and *Wettinia hirsuta* were modeled following the same techniques used for stems and branches, making use of field notes, photographs, and reconstruction of vector files to complete the matrix of areas and volumes (Fig. 5).

**Classification of tree models based on architecture**

There are several ways to classify tree architecture, including growth characteristics, main branching, bifurcation, apical branch growth, location of growth meristems, types of secondary branching, leaf arrangement, and angles of leaf distribution along the branches. Halle and Oldeman (1970) proposed a system that grouped these characteristics into general architectural models and highlighted the main features of structural and spatial configuration acquired by different tropical species. To adjust the characteristics of some types of branching in three-dimensional models, while taking into account possible causes of variability in biomass estimates, the classification proposed by Keller (2004) was used. This approach incorporated and improved the description of the architectural models of Halle and Oldeman (1970) and helped resolve a considerable amount of ambiguities presented during the fieldwork.

For tree architecture classification, a matrix was built that included information on overall growth pattern, architectural group, and principal features. Species that generally showed monopodial growth, that is, those in which a leading stem or principal growth axis (hypersilepsia) was detected, as compared with lateral meristems, were given a value of 1 (one). Species with sympodial growth patterns, in which an apical meristem was replaced by a lateral one or where several leader branches were identified, were given a value of 0 (zero). Branching patterns seen in monopodial trunks were classified according to the types listed in Appendix S1 (Keller 2004).

**Sample acquisition for carbon analysis**

To determine carbon content in the four vegetation communities analyzed, all woody individuals with DBH (diameter at breast height) greater than 5 cm were counted within 500-$m^2$ plots. In addition to the typical structural information (DBH, height, and coverage) for each individual, organ samples were taken from each to be used for carbon analysis. The stem samples were obtained by extracting a small core an inch in diameter, with depths ranging from 1 to 5 cm (without bark). The depth depended on thickness of the trunk, as well as what was determined to cause the least harm to the individual. For small individuals, the sample never exceeded 1 inch in depth and 1 inch in diameter. In these cases, additional bark was removed by scraping an area of up to 20 $cm^2$ and was added to the respective sample. For the more-numerous organs, such as leaves and branches, three samples per organ were taken to obtain replicates, the number determined by the cost of the analysis. To collect branch and leaf samples from larger individuals, transects of 100 $m^2$ were set up, or the tissues were acquired from sites with similar physical conditions. Samples were weighed, both fresh and dry (dried at 80° C), and volume was calculated by displacement of water in a micrograduated test tube. A total of 2,232 pulverized samples weighing 0.02 g each from 186 individuals were dried again at 105 ° C, then processed in an automatic analyzer (LECO CHN-600) (Vásquez & Arellano 2012, Lozano 2012, Sheldrick 1984).

With the LECO analyzer, total carbon content was quantified by combustion of the tissues at 950° C (Vásquez & Arellano 2012, Lozano 2012, Sheldrick 1984).The analyzer's sensitivity of 0.01% is of great importance to the regional assessment of carbon content in the different vegetation communities. Also, because it is generated by a calibrated quantification method, it differs substantially from other methods such as LOI (Loss On Ignition), which not only requires bulk samples but introduces significant error by relying on the various weighing methods used by researchers. To be employed properly, the LOI method requires an analytical balance and the collection of several replicates of large samples to generate the calibration curves by species. To obtain data for the LECO analyzer, only 0.02 g of a sample is required, as well as monitoring every 10 samples with low-cost, standard organic targets with known carbon content, such as ethylenediaminetetraacetic acid (EDTA), sucrose, and leucine, in order to make small adjustments if necessary to keep the process calibrated (Vásquez & Arellano 2012, Lozano 2012, Sheldrick 1984).

**Comparisons with other methodologies**

Vegetation structure and the variation in biomass estimates by traditional methods were assessed in Vásquez & Arellano (2012), on whose results this study is partly based. The data were explored using descriptive statistical analysis. Frequency distributions, central tendency measurements, and biomass distribution were evaluated by means of the basal area, density (number of elements), and height of individuals belonging to 12 types of forest, among which the following associations previously described by Rangel *et al.* (2010) and Avella and Rangel (2011) are the most important in terms of area: Jacarando copaiae–Pouterietum multiflorae (superhumid), Protio aracouchini–Viroletum elongatae (very humid), Trichilio hirtae–Schizolobietum parahibi (humid), and the secondary community dominated by *Acalypha diversifolia* and *Guazuma ulmifolia* (semihumid). Also, results from different models were compared in order to choose the allometric model providing the best fit between DBH, height, and specific weight of the wood density. The Chave *et al.* models (2005) were chosen because, in addition to presenting a good apparent fit, they evaluated the DBH intervals between 5 and 156 cm and estimated biomass in tropical wet and dry forests. In Colombia, Alvarez et al. (2012) published some edited versions of the Chave *et al.* (2005) equations, which showed a similar trend and increased the biomass estimates of those formulas.

To compare the results found in this study (by integrating three-dimensional modeling, numerical integration, voxelization, and data from small samples—ca. 16.39 $cm^3$ per organ—collected in the field) with biomass estimates from traditional allometric methods, a comparison of covariance was performed. This analysis assumed that the biomass matrix found in this study (**X**) corresponded to a simple random sample of size $n_x$, from a normal multivariate law $X \sim \mathrm{N_m}\left(\mu_X, \sum_X\right)$, and that the matrix **Y** (biomass matrix found by the allometric equations) corresponded to another simple random sample independent from the previous one, of size $n_y$, from a multivariate law $Y \sim \mathrm{N_m}\left(\mu_Y, \sum_Y\right)$. Equality of covariance was evaluated by the contrast of the following hypothesis (Peña 2002): $H_0 : \sum_X = \sum_Y = \sum$ . The contrast probability ratio was used in this case, which was found by Equation 2.

Equation 2. $\lambda_R = \dfrac{|S_X|^{\frac{n_X}{2}} |S_Y|^{\frac{n_y}{2}}}{|S|^{\frac{n}{2}}}$ ,

where *Sx* and *Sy* represented the sampled covariance matrices in each group, $n = n_x + n_y$, and *S* was the common covariance matrix, found by weight (Peña 2002). Under the null hypothesis it was given that $-2\log(\lambda_R) \sim \chi^2_{\mathrm{q}}$, where $\mathrm{q} = \dfrac{(\mathrm{g}-1)p(p+1)}{2}$, taking into account that $-2\log(\lambda_R) = n * \log|S| - \left(n_X \log|S_X| + n_Y \log|S_Y|\right)$ (Leigh 1999).

Thus, the null hypothesis was rejected if the value of this statistic belongs to the critical region $[x_{1-\alpha}, +\infty)$, where $x_{1-\alpha}$ represents the percentile-α (1-α)100%. When the null hypothesis of the equal covariance test was not rejected, we proceeded to perform a comparison of means by contrasting null hypothesis $H_0 : \mu_X = \mu_Y$ , using the statistic based on the $\Lambda$ distribution of Wilks, $\Lambda(p, n-g, g-1)$. The statistic was found by the formula $\Lambda = \frac{|W|}{|B+W|} = \frac{|W|}{|T|}$, where $W = n_X S_X + n_Y S_Y$ was the scattering matrix within groups and

$B = n_X (\bar{X} - \bar{Z}),(\bar{X} - \bar{Z}) + n_Y (\bar{Y} - \bar{Z}),(\bar{Y} - \bar{Z})$ was the scattering matrix between groups. $T = W + B$ was considered the total scattering matrix, which contained the global mean vector $\bar{Z} = \frac{n_X \bar{X} + n_Y \bar{Y}}{n}$ (Peña 2002).

## RESULTS

### A biomass comparison using allometric methods

Biomass was determined using three-dimensional models of the species and tissue density values (see Methods). A summary of estimated volume and biomass values for individuals in the plots is given in Table 1. Total values are shown in Tables S1–S3. These results were compared with those of the most-accurate traditional methodology based on allometric equations for the various organs analyzed stems, large branches, and twigs (Overman *et al.* 1990). The hypothesis of equality of covariances for the X matrix (DBH; height; biomass of the stems, main branches, and secondary branches or twigs; leaf biomass; and total biomass using three-dimensional modeling) and Y matrix (the same structural variables as X, plus the biomass of the organs estimated with allometric equations) (Overman *et al.* 1990) was rejected in all cases. The rejection of the hypothesis indicated the important differences between the methods (Table 2).

This outcome highlighted the bias observed over the full range of DBH values, because the allometric formulas were unable to incorporate information about the heterogeneity of the individuals' shapes. Despite this demonstrated bias, allometric equations are widely used to estimate biomass in tropical forests (Chave *et al.* 2005). Vásquez and Arellano (2012) and Lozano (2012) noted their similarities to nine other known methodologies. These equations not only present the highest correlations among the resulting data but also are able to integrate wood density records. It is noteworthy that Álvarez *et al.* (2012) published several equations that, using the variables DBH, height, and wood density, generate a greater biomass estimate, although describing the same mathematical behavior shown by the equations of Chave *et al.* (2005).

Figure 6 shows a comparison of estimated biomass observed for forests on the gradient from superhumid to humid climates dominated by the associations *Jacaranda copaia*—*Pouteria multiflora*, *Protium aracouchini*—*Virola elongata,* and *Trichilia hirta*—*Schizolobium parahyba,* using the methodology proposed in this study and the traditional methodology to analyze individuals with DBH between 1 and 20 cm. For some individuals in this range, the biomass estimates found in this study are similar to those generated by the allometric equations of Overman *et al.* (1994). For others the biomass values behave as outliers around the values from the allometric method, that is, show higher variability, which indicates that this study captured more of the heterogeneity present in the samples. The three-dimensional technique incorporates the high structural variability present in the sampled regions and produces estimates of the biomass of individuals with greater accuracy. In general, results obtained by traditional methods, both for estimates of organs (Overman *et al.* 1994; Overman *et al*. 1990) and complete individuals (Chave 2005), tend to homogenize the records.

Similar tendencies to the ones previously given are presented for comparative records of main branching, secondary branches (twigs), leaves, and biomass for the whole individual (Figure 6). Table S2a shows biomass estimates for individuals in humid to superhumid climates with DBH of 1–20 cm. Differences can be noted between these results and those found with the equations of Overman *et al.* (1994).

Figure 7 shows the comparison of methods for samples with DBH between 20 and 200 cm. It is important to mention that since biomass increases are on a cubic scale because of the three-dimensionality of the measurement, differences between the methods increase similarly when the DBH is increased. Overman *et al.* (1994) recognized many deficiencies in their method, which are reflected in the concentration of values obtained by this technique in the center of the figure (Figure 7). Records of total aboveground biomass (the sum of all components per individual) behave similarly to those of the main branching (main stem and branches), more evidence of the large contribution of this component to the total. Table S2b shows the total values for biomass estimates of samples with DBH of 20–200 cm. For vegetation types in semihumid climates the comparisons are given in Table S2.

When basic structural data (DBH, height, and density) were explored using the most common allometric equations (Vásquez & Arellano 2012), the results of nine equations were found to be highly correlated, evidence of the widespread use of this methodology. The equations of Chave *et al.* (2005) were chosen as representative of the others. These equations were compared with biomass records estimated by our three-dimensional modeling methodology. When values generated by the equations of Chave *et al.* (2005) and Overman *et al*. (1994) were compared with those generated by our methods, the former appeared to be more concentrated in the smaller DBH range and the latter in the larger range (these results can be observed in Tables S2 and S3).

Results of the neural network of the estimated carbon content for more than 6,000 individuals of different species belonging to 27 vegetation types in the studied area, are compared with the values from the allometric method (Chave et al. 2005). These are shown in Figure 8.

Carbon per hectare was estimated on the basis of results of the spatial distribution of vegetation types and covers generated by the new CCA–Fuzzy Land Cover methodology (Arellano & Rangel 2015, in press). Values were extrapolated and calibrated using density analysis and artificial back-propagation neural networks for the surface sampled (Appendix S2). A total of 9,559.64 Mg C was recorded for 54 vegetation types (average, 177.03 Mg C $ha^{-1}$) with the new methodology; total carbon estimated using traditional methods was 3,316.08 Mg C (average, 61.41 Mg C $ha^{-1}$). The differences between the two methods was 6,243.56 Mg C for the 54 vegetation types studied and 115.62 Mg C $ha^{-1}$ for the averages, or 65.3% of the result from nondestructive methods. The highest carbon content per hectare was found in primary tropical forests in superhumid climates in sites associated with bodies of water, with little or no intervention, and dominated by *Macrolobium ischnocalyx* and *Peltogyne purpurea*: 523.4 Mg C (traditional methodology, 290.63 Mg C $ha^{-1}$; a difference of 232.77 Mg C $ha^{-1}$, or 44.47%). The second highest carbon content per hectare was in primary tropical forests in superhumid climates, with low to medium intervention, and dominated by *Jacaranda copaia* and *Pouteria multiflora*: 437.25 Mg C $ha^{-1}$ (traditional methodology, 196.01 Mg C $ha^{-1}$; a difference of 241.24 Mg C $ha^{-1}$, or 55.17%). The third highest was in primary tropical forests in superhumid climates in flooded sites, with little or no intervention, and dominated by *Prestoea decurrens* and *Trichilia poeppigii*: 428.62 Mg C $ha^{-1}$ (traditional methodology, 267.71 Mg C $ha^{-1}$; a difference of 160.91 Mg C $ha^{-1}$, or 37.54%). The results per hectare (Appendix S2) represent information comparable to nearly the entire carbon storage gradient reported in the pantropical map published in 2012 (Baccini 2012). Information from the map of tree-height distribution (Simard *et al.* 2011) was used to generate a calibration model for the artificial neural network and to create a calibrated pantropical carbon map (for the period 2007–2008; Fig. 9, Table S3). It is worth mentioning that the differences reported by Mitchard *et al*. (2014) for the 2012 pantropical carbon-storage map (Baccini *et al.* 2012) do not arise from regional variations in wood density, and its relation to height and DBH. Another approach reported by Avitabile *et al.* (2016) results in estimates of aboveground biomass around 18% lower than those of Baccini *et al.* (2012). This calibration was developed using higher-quality field data but has the same source of bias as the allometric equations.

Training values used in the artificial neural network for the study area were between the minimum and maximum limits of 0 and 190 Mg C $ha^{-1}$ and 0 and 38 m for the variable of individual height, according to Baccini *et al.* (2012) and Simard *et al*. (2011), respectively. The values in the response range, 0 to 523 Mg C $ha^{-1}$, were based on carbon data for the region derived from this report. The results of the processes of training, testing, and validation had correlation coefficients R = 0.996, 0.997, and 0.988, respectively (overall, R = 0.995), which legitimized this process (Fig. 10). Using these methods we thus determined that the maximum carbon storage in aboveground biomass in the tropics is approximately 605 Mg C $ha^{-1}$ (251.5 Mg C $ha^{-1}$, Baccini *et al.* 2012). In contrast, using the calibration model the carbon stored in the pantropical region would be around 723.97 Pg C (Fig. 9). Compared with the results of our methodology, the estimates of Baccini *et al.* (2012) and Feldpausch *et al.* (2012) represent only 31.59% (228.7 Pg C) and 39.38% (285.1 Pg C), respectively, of the carbon stored in this region. These large differences can be attributed to the fact that high-precision data derived by Light Detection and Ranging (LIDAR) used in the first map (Baccini *et al*. 2012) were calibrated with carbon values per unit area, which were derived from half the biomass estimates of sampled individuals using allometric formulas. The most complete formulas (Chave *et al.* 2014, Chave *et al.* 2005, Sileshi 2014, Rutishauser *et al.* 2013) incorporate diameter, height, and wood density. However, they do not consider the variation in carbon stored in different organs or different species, or information on the wide variety of tree architectures present in different vegetation communities.

Table 3 shows an arbitrary comparison among various ranges of carbon storage per hectare in the pantropical region of the world (excluding the Australian tropics). These values are the totals of calibrated carbon measurements per pixel from the matrix arrangements on the maps. Additionally, the summation of data from the pantropical map published in 2012 (Baccini *et al.* 2012) and available from the Woods Hole Research Center (2013) does not correspond to the final value reported in Baccini *et al*. (2012) of 228.7 Pg C, because the value, 227.42 Pg C, was the result of the information available from the internet source. Despite this adjustment, comparisons, trends, and differences are valid for the analyzed regions. As was expected, a huge difference (between -405.80% and 100%) in the classes evaluated was observed in this redistribution of carbon content values for pantropic regions. Therefore, three new classes between 300 and 605 Mg C ha$^{-1}$ were generated, which sequester 583.55 Pg C, that is, slightly more than 80% of the total pantropical value. The largest variation was found for the regions with a content of between 1 and 50 Mg C ha$^{-1}$, which store barely 8.83 Pg C (44.68 Pg C, -405.8%; Baccini *et al*. 2012). Thus, we estimated that America stores in its aboveground biomass about 382.85 Pg C (117.7 Pg C, 30.74%; Baccini *et al.* 2012), followed by Africa with 175.77 Pg C (64.5 Pg C, 36 69%; Feldpausch *et al.* 2012) and Asia-Oceania with 165.35 Pg C (46.5 Pg C, 28.12%; Baccini *et al.* 2012). Table 4 shows the values of countries with the highest carbon stocks.

**Validation of the estimation methods**

The validation process consisted of first measuring the volumes of 10 individuals of *Aphelandra pulcherrima* (with heights from 1.54 to 2.9 m) by cutting, weighing, and calculating green volume by water displacement using test tube (Fig. 11). The mean volume by water displacement, 1,240 $cm^3$, was compared with the mean volumes measured by numerical integration (1,217.76 $cm^3$, 1.798% difference), by voxelization with 1-cm resolution (1,211 $cm^3$, 2.338% difference), and by voxelization with 1-mm resolution (1,189.92 $cm^3$, 4.038% difference).

**Comparison of voxelization methods and numerical integration**

To determine the most sensitive volumetric classification method with regard to calibrated values (in $cm^3$), overall volume estimates acquired through voxelization and uncalibrated numerical integration were evaluated in a Cohen's kappa reliability analysis and an ROC quality analysis (Cardillo 2007). For the former, a typical confusion matrix was generated without modifying the values. For the latter, the proximity of raw results between 0 (zero) and 800 $cm^3$ to the calibrated value was considered positive (labeled 0 [zero], true negatives), and all records not found within these limits were considered negative (labeled 1 [one], false positives). As was discussed in the methodology, in the voxelization calculations different cubic units were employed; thus, as with technologies based on pixel analysis (Raster), the best results can be expected from datasets with high density (resolution), which in this case generally correspond to models limited to 15 m in length.

Test results showed the percentage contributed by the two methodologies to the uncalibrated volume estimate, and within these results (volumes), which sectors are better represented by one or another methodology (without calibration). It is important to note that the records that helped most in correcting inconsistencies caused by the choice of different measurement units mostly belonged to estimates in cubic centimeters for secondary branches of the various species modeled.

Voxelization produced the record with the highest concordance strength, with kappa = 0.6315 (kappa = 0.4342 using numerical integration), whereas in the ROC quality analysis, the highest record was produced using numerical integration to estimate volume, with an area under the curve of 0.9560, and a confidence interval (CI) of 0.9285–0.9834 (area under the ROC curve using voxelization = 0.91447, CI 0.8667–0.9623) (Appendix S3). As shown in Fig. 12, it appears that the quality of estimates with numerical integration exceeded that of voxelization because it generated a lower rate of false positives and greater specificity for the evaluated volumes. However, these results should be reinterpreted on the basis of the arguments shown throughout this study and the noted benefits of the voxelization method. The ROC method evaluated a wide range of volumes (between 82 and 155,116,759 $cm^3$), but because of technological limitations, voxelization produced higher-quality results with smaller volumes (between 82 and 2,403,575 $cm^3$), and therefore the range of accuracy of this method is concentrated on the lowest rates of true positives. This boosts the quality of the numerical integration method, because due to its vector characteristics it is not limited by height or volume. It is important to emphasize that the numerical integration method showed minimal differences as compared with voxelization when two factors were present: the tree was represented by a homogeneous mesh model for all modeled organs (vector structure of vertices, normals, and faces), and voxelization was performed with cubic centimeter units. This leads to the conclusion that homogeneous mesh models allow adequate volume estimates with the numerical integration method, but this is limited by the available computing capacity.

## Discussion

These results suggest that the voxelization method is quite accurate, even more than the water displacement method for calculating volume, due to the fact that the sample (a whole individual) loses its integrity when it is cut in multiple pieces, and its volume tends to decrease when observed at more-precise resolutions of less than 1 $cm^3$.

The wide range of estimates of carbon reserves, together with the factors outlined above, indicates that data from remote high-precision sensing methodology (LIDAR) are negatively affected by the need to include field data in calibrations and the fact that they are typically derived from the traditional destructive methodologies described above. Moreover, it is problematic that results generated to date do not take into account types of vegetation growing in different soils; thus the outcomes are highly biased because these studies do not reflect the variation in pantropical geography. In addition, if floristic composition and vegetation assemblages are ignored when constructing these thematic maps, final estimates will include unpredictable increases in error, regardless of the quality and resolution of inputs used in their construction. Some authors use generic logarithmic equations to estimate error in the underestimation of biomass stocks, which usually ranges between 0% and 40% for certain regions of the pantropics like East Kalimantan (Rutishauser *et al.* 2013). However, according to our results, the distribution error associated with the traditional method may vary between 36% and 96%, without taking into account the error from imprecise coverage maps (Appendix S2).

When we analyzed values for the percentage of greenhouse gases (GHG) attributable to pantropical deforestation as reported by the IPCC and others in 1996 (10.49%), 2000 (18.3%), 2005 (11.3%), and 2010 (10.3%) (IPCC 1996, Herzog & Timothy 2009-2000, ECOFYS 2013), we concluded that the contribution to GHG from pantropical deforestation would be between 26.15% and 32.6% based solely on our carbon estimate, without taking into account data from maps of deforestation and degradation coverage (Arellano 2012). This establishes deforestation as the principal cause of climate change. In coming years, we will see one of three possible scenarios for GHG levels. The first involves GHG totals higher than current measurements, which would result from GHG from other sources being accurately measured. The second scenario shows GHG totals remaining constant, implying that other sources of GHG are poorly measured and are lower than currently reported values. And the third, most-complex scenario implies a lack of knowledge of other variables and unknown GHG totals levels.

The methodology presented here can improve knowledge of distribution of biomass and carbon on the planet and is applicable to projects that characterize and evaluate the environmental services of biodiversity, such as carbon storage for REDD+ projects.

## Endnotes

### Supplementary Information

### Acknowledgments

Part of this work was financed by the Bicentennial Project of the National University of Colombia, and support from the Colombian Geological Service for elemental tissue sample analysis was also received. We thank Adela Lozano for her active participation in various phases of the project and Andres Avella for his contributions. The reviews and valuable comments by the Gerardo Aymard, Gustavo Romero and Susan Donoghue. Also the collaboration of Laura Pérez, Vanessa Pérez, Victor Chavez and Henry Arellano-C.

## Author contributions

The study idea, research design, aboveground carbon sampling design, development methodology, data analysis, and preparation of figures for the main article and supplementary information were provided by H.A.P.

O.R.C. was responsible for the collection and preparation of structural data, for conducting the local phytosociological analysis that supported this study, and for providing logistic and field support for carrying out the research.

H.A.P. wrote the paper and interpreted the results, and both authors reviewed the manuscript.

## Author information

The authors declare no competing financial interest; details accompany the full-text HTML version of the paper at www.colombiadiversidadbiotica.com. There is no commercial affiliation between the authors and the entities referred to herein; thus, no commercial use of the information presented in this document without the consent of the principal author is allowed.  Correspondence and requests for materials should be addressed to harellano@unal.edu.co or harellano@gmail.com

# Table 1 | Summary of volume and biomass of individuals in plots.

| ASSOCIATION | SYMBOL MAPPING (13b) | SPECIES | DBH (cm) | HEIGHT (m) | TOTAL VOLUME LEAVES (cm3) | TOTAL ABOVEGROUND BIOMASS (Kg) | TOTAL ABOVEGROUND BIOMASS EQUATIONS OVERMAN et. al. (1990) (Kg) | TOTAL ABOVEGROUND BIOMASS EQUATIONS CHAVE et al. (2005) (kg) |
|---|---|---|---|---|---|---|---|---|
| Jacarando copaiae - Pouterietum multiflorae | BHtf2/Jco-Pmu | *Amphirrhox longifolia* | 7.92 | 6.04 | 685.05 | 4.59 | 13.83 | 11.16 |
| | | *Brosimum utile* | 9.20 | 6.24 | 5597.63 | 4.67 | 20.24 | 17.19 |
| | | *Crepidospermum rhoifolium* | 5.49 | 7.84 | 2781.03 | 8.08 | 6.57 | 4.69 |
| | | *Dendrobangia boliviana* | 29.87 | 21.06 | 357970.05 | 1245.04 | 527.42 | 559.05 |
| | | *Dendrobangia boliviana* | 40.23 | 19.26 | 192102.28 | 2364.19 | 870.11 | 927.64 |
| | | *Dialium guianense* | 69.49 | 30.53 | 11151.79 | 25823.46 | 6240.60 | 6375.30 |
| | | *Dipteryx oleifera* | 81.69 | 30.29 | 95601.20 | 26112.26 | 7804.09 | 7882.48 |
| | | *Dipteryx oleifera* | 163.92 | 36.42 | 130032.75 | 52314.84 | 42987.46 | 38161.36 |
| | | *Duguetia flagellaris* | 12.44 | 15.54 | 1281.86 | 78.94 | 106.07 | 105.23 |
| | | *Eschweilera pittieri* | 29.87 | 19.94 | 26882.91 | 1061.47 | 678.76 | 722.18 |
| | | *Guarea gomma* | 32.92 | 18.98 | 153538.76 | 454.17 | 466.06 | 492.79 |
| | | *Guarea gomma* | 43.89 | 17.03 | 722495.27 | 3043.05 | 738.24 | 786.15 |
| | | *Guarea sp 2* | 3.47 | 4.35 | 10777.23 | 1.18 | 2.71 | 1.60 |
| | | *Helianthostylis sprucei* | 12.19 | 6.89 | 1040.06 | 55.35 | 34.87 | 31.51 |
| | | *Helicostylis tomentosa* | 3.66 | 5.41 | 10805.71 | 4.59 | 3.69 | 15.87 |
| | | *Helicostylis tomentosa* | 9.14 | 5.86 | 29665.75 | 11.38 | 18.87 | 2.35 |
| | | *Inga pilosula* | 14.02 | 15.12 | 8992.04 | 143.14 | 72.55 | 70.05 |
| | | *Iryanthera hostmannii* | 6.10 | 3.05 | 464.49 | 1.01 | 5.48 | 3.78 |
| | | *Iryanthera hostmannii* | 13.41 | 25.56 | 70792.74 | 1067.84 | 151.23 | 153.25 |
| | | *Iryanthera hostmannii* | 27.98 | 18.17 | 57268.97 | 1276.78 | 384.19 | 404.31 |
| | | *Jacaranda copaia* | 37.19 | 25.60 | 2253.29 | 519.43 | 771.59 | 821.98 |
| | | *Lindackeria laurina* | 27.43 | 20.79 | 26941.27 | 1413.18 | 421.62 | 444.76 |
| | | *Micropholis guyanensis* | 39.01 | 21.16 | 25546.06 | 1145.76 | 1014.30 | 1081.81 |
| | | *Nectandra membranacea* | 1.83 | 2.01 | 1091.46 | 54.38 | 0.50 | 0.17 |
| | | *Pentaclethra macroloba* | 48.16 | 23.83 | 18330.91 | 5153.94 | 1875.04 | 1990.46 |
| | | *Pouteria multiflora* | 70.10 | 27.96 | 41809.83 | 7467.86 | 2743.95 | 2889.67 |
| | | *Protium aracouchini* | 11.03 | 17.42 | 483127.56 | 213.15 | 65.71 | 62.96 |
| | | *Protium aracouchini* | 34.75 | 20.90 | 109084.27 | 1658.91 | 703.81 | 749.14 |
| | | *Schefflera morototoni* | 15.85 | 23.22 | 20.87 | 179.00 | 146.74 | 148.45 |
| | | *Sorocea trophoides* | 23.16 | 23.46 | 65211.60 | 731.90 | 374.13 | 393.43 |
| | | *Sterculia apetala* | 16.46 | 12.09 | 237988.14 | 83.33 | 58.46 | 55.47 |
| | | *Swartzia robiniifolia* | 16.46 | 15.04 | 5258.29 | 603.04 | 178.61 | 182.56 |
| | | *Theobroma glaucum* | 18.29 | 10.98 | 21564.46 | 33.89 | 118.91 | 118.83 |
| | | *Theobroma glaucum* | 20.73 | 21.09 | 682692.48 | 299.09 | 281.23 | 293.05 |
| | | *Trichilia sp 1* | 2.99 | 3.00 | 171.61 | 0.11 | 1.59 | 0.82 |
| | | *Virola elongata* | 21.34 | 14.03 | 2678.83 | 261.18 | 154.42 | 156.66 |
| | | *Virola elongata* | 121.92 | 13.80 | 454611.24 | 607.98 | 5702.06 | 5849.72 |
| Protio aracouchini - Viroletum elongatae | BHmhtf1/Par-Vel | *Amphirrhox longifolia* | 10.97 | 7.45 | 723.84 | 42.86 | 29.67 | 26.35 |
| | | *Amphirrhox longifolia* | 14.02 | 11.95 | 3454.47 | 235.00 | 71.60 | 69.07 |
| | | *Brosimum guianense* | 38.40 | 24.45 | 25293.61 | 3192.35 | 2916.83 | 1323.35 |
| | | *Brosimum guianense* | 58.52 | 24.40 | 44034.98 | 5394.16 | 1241.30 | 3066.69 |
| | | *Brosimum utile* | 13.41 | 10.44 | 7889.47 | 22.22 | 63.91 | 61.09 |
| | | *Carapa guianensis* | 7.92 | 14.01 | 169347.98 | 125.57 | 21.85 | 18.73 |
| | | *Cariniana pyriformis* | 57.30 | 30.06 | 3107.33 | 7700.54 | 2585.82 | 2727.23 |
| | | *Cariniana pyriformis* | 68.88 | 19.85 | 3632.20 | 2250.09 | 2464.89 | 2602.65 |
| | | *Caryocar amygdaliferum* | 23.77 | 6.39 | 39998.21 | 5.45 | 19832.29 | 106.60 |
| | | *Caryocar amygdaliferum* | 123.75 | 41.71 | 423433.20 | 30362.98 | 107.36 | 18838.25 |
| | | *Castilla elastica* | 8.53 | 7.97 | 33418.48 | 7.02 | 13.48 | 10.83 |
| | | *Cavanillesia platanifolia* | 79.86 | 37.03 | 329884.40 | 15519.07 | 5208.23 | 5364.57 |
| | | *Cecropia sp 2* | 5.49 | 7.90 | 172.26 | 1.66 | 8.05 | 5.96 |
| | | *Compsoneura mutisii* | 14.02 | 7.06 | 4504.31 | 8.49 | 28.98 | 25.67 |

| ASSOCIATION | SYMBOL MAPPING (13b) | SPECIES | DBH (cm) | HEIGHT (m) | TOTAL VOLUME LEAVES (cm3) | TOTAL ABOVEGROUND BIOMASS (Kg) | TOTAL ABOVEGROUND BIOMASS EQUATIONS OVERMAN et. al. (1990) (Kg) | TOTAL ABOVEGROUND BIOMASS EQUATIONS CHAVE et al. (2005) (kg) |
|---|---|---|---|---|---|---|---|---|
| Protio aracouchini - Viroletum elongatae | BHmhtf1/Par-Vel | *Protium aracouchini* | 39.75 | 19.76 | 119038.86 | 1332.18 | 825.79 | 880.14 |
| | | *Pseudolmedia laevigata* | 24.38 | 12.16 | 41774.76 | 441.08 | 219.13 | 226.10 |
| | | *Pterocarpus rohrii* | 9.14 | 7.48 | 296640.81 | 35.51 | 16.81 | 13.93 |
| | | *Qualea aff. dinizii* fitogeográfico! | 37.19 | 18.43 | 21082.39 | 1704.02 | 666.05 | 708.50 |
| | | *Schefflera morototoni* | 26.21 | 22.07 | 18.98 | 426.68 | 367.16 | 385.90 |
| | | *Schizolobium parahyba* | 32.92 | 19.16 | 74991.47 | 530.83 | 613.24 | 651.62 |
| | | *Simaba cedron* | 16.46 | 10.27 | 5746.67 | 63.97 | 58.27 | 55.28 |
| | | *Sloanea brevispina* | 26.82 | 16.18 | 56436.79 | 871.97 | 387.34 | 407.71 |
| | | *Sorocea trophoides* | 13.41 | 12.44 | 5065.13 | 31.64 | 72.42 | 69.93 |
| | | *Swartzia simplex* | 12.80 | 9.06 | 4424.83 | 74.82 | 69.17 | 66.54 |
| | | *Tetragastris panamensis* | 18.90 | 12.93 | 76500.16 | 183.20 | 151.91 | 153.98 |
| | | *Theobroma glaucum* | 3.66 | 2.84 | 19776.82 | 1.39 | 2.15 | 1.20 |
| | | *Virola elongata* | 4.88 | 4.19 | 2080.79 | 4.88 | 4.32 | 12.79 |
| | | *Virola elongata* | 7.92 | 7.14 | 911.72 | 0.53 | 15.60 | 2.84 |
| | | *Virola elongata* | 14.63 | 13.37 | 2359.45 | 115.74 | 83.62 | 81.61 |
| | | *Virola reidii* | 5.49 | 9.17 | 8592.20 | 11.06 | 8.10 | 6.00 |
| | | *Virola reidii* | 111.56 | 29.09 | 377991.33 | 10297.80 | 6997.34 | 7108.09 |
| | | *Virola sebifera* | 84.12 | 26.98 | 40686.00 | 8427.72 | 5032.68 | 5191.34 |
| Trichilio hirtae-Schizolobietum parahibi | Bhtf1/Thi-Spa | *Ampelocera macrocarpa* | 9.14 | 9.16 | 2239.63 | 9.33 | 32.59 | 29.25 |
| | | *Ampelocera macrocarpa* | 29.87 | 19.79 | 14811.21 | 1113.50 | 634.10 | 674.10 |
| | | *Anacardium excelsum* | 38.40 | 18.95 | 15060.63 | 739.64 | 612.09 | 650.38 |
| | | *Anacardium excelsum* | 61.20 | 11.36 | 844.87 | 150.64 | 928.43 | 990.06 |
| | | *Annona spraguei* | 34.75 | 19.54 | 150966.20 | 810.69 | 620.26 | 659.18 |
| | | *Aspidosperma curranii* | 35.97 | 18.00 | 53409.16 | 1370.10 | 706.63 | 752.16 |
| | | *Astronium graveolens* | 22.56 | 13.29 | 37229.57 | 907.67 | 207.24 | 213.31 |
| | | *Pachira quinata* | 28.04 | 14.78 | 90556.71 | 666.42 | 221.01 | 228.12 |
| | | *Bursera simaruba* | 39.62 | 21.19 | 471492.64 | 281.76 | 579.94 | 615.72 |
| | | *Bursera simaruba* | 90.22 | 25.94 | 423216.08 | 4017.15 | 3745.98 | 3907.99 |
| | | *Caesalpinia glabrata* | 42.67 | 16.13 | 3022.12 | 1488.85 | 1157.50 | 1234.36 |
| | | *Castilla elastica* | 33.53 | 20.23 | 133429.23 | 775.69 | 402.47 | 424.06 |
| | | *Ceiba pentandra* | 39.62 | 17.38 | 53443.27 | 135.64 | 394.10 | 415.01 |
| | | *Ceiba pentandra* | 197.51 | 40.44 | 195525.38 | 54351.06 | 25807.52 | 23995.74 |
| | | *Chrysophyllum lucentifolium* | 9.75 | 10.08 | 45286.54 | 19.65 | 37.59 | 34.23 |
| | | *Chrysophyllum lucentifolium* | 34.75 | 19.61 | 37632.68 | 908.81 | 793.45 | 845.44 |
| | | *Crateva tapia* | 17.07 | 14.08 | 33614.86 | 35.59 | 112.86 | 112.42 |
| | | *Gyrocarpus americanus* | 9.14 | 5.15 | 23737.54 | 12.62 | 6.25 | 4.42 |
| | | *Hura crepitans* | 176.78 | 47.92 | 238866.63 | 35121.83 | 36014.38 | 32512.16 |
| | | *Machaerium goudotii* | 70.10 | 17.44 | 697.76 | 4182.66 | 3035.85 | 3188.23 |
| | | *Marila laxiflora* | 30.48 | 24.14 | 67911.25 | 612.44 | 447.21 | 472.42 |
| | | *Pentaplaris doroteae* | 10.36 | 10.38 | 12140.02 | 22.40 | 34.57 | 117.07 |
| | | *Pentaplaris doroteae* | 17.68 | 13.38 | 8707.46 | 8.40 | 117.25 | 31.21 |
| | | *Pentaplaris doroteae* | 90.22 | 28.04 | 404690.93 | 5726.41 | 6254.27 | 6388.59 |
| | | *Pouteria subrotata* | 9.75 | 9.80 | 47343.19 | 38.12 | 36.08 | 32.72 |
| | | *Pouteria subrotata* | 21.95 | 11.53 | 61628.61 | 251.05 | 190.20 | 194.99 |
| | | *Pseudobombax septenatum* | 120.70 | 29.79 | 234312.72 | 16874.63 | 7513.78 | 7604.60 |
| | | *Toxicodendron striatum* | 15.24 | 7.41 | 39148.21 | 136.58 | 54.14 | 51.03 |
| | | *Schizolobium parahyba* | 41.45 | 19.63 | 79315.97 | 842.11 | 807.01 | 860.00 |
| | | *Schizolobium parahyba* | 60.35 | 27.63 | 62128.17 | 3573.41 | 2429.61 | 2566.26 |
| | | *Aralia excelsa* | 45.72 | 15.70 | 360.59 | 831.51 | 642.73 | 683.39 |
| | | *Trichilia hirta* | 9.75 | 6.52 | 10888.47 | 26.08 | 25.34 | 22.11 |
| | | *Trichilia hirta* | 54.25 | 26.88 | 14528.02 | 4379.04 | 2284.67 | 2416.44 |

| | | | | | | |
|---|---|---|---|---|---|---|
| *Compsoneura mutisii* | 15.24 | 12.34 | 4522.96 | 45.92 | 56.08 | 53.02 |
| *Compsoneura mutisii* | 28.65 | 17.04 | 2369.47 | 384.79 | 249.62 | 258.95 |
| *Couratari guianensis* | 7.32 | 8.78 | 62703.71 | 75.02 | 14.21 | 11.51 |
| *Croton pachypodus* | 16.46 | 9.96 | 14999.10 | 126.59 | 86.20 | 84.32 |
| *Croton pachypodus* | 17.68 | 14.60 | 25003.89 | 390.63 | 141.21 | 142.55 |
| *Dendrobangia boliviana* | 7.92 | 6.57 | 136885.61 | 11.22 | 13.58 | 26.60 |
| *Dendrobangia boliviana* | 9.14 | 8.07 | 71069.06 | 20.14 | 19.53 | 16.51 |
| *Dendrobangia boliviana* | 11.58 | 7.49 | 103435.07 | 9.88 | 29.92 | 10.92 |
| *Dendropanax arboreus* | 12.80 | 9.30 | 1316.98 | 32.09 | 42.39 | 39.07 |
| *Dialium guianense* | 41.45 | 20.39 | 14088.14 | 1606.32 | 1186.20 | 1264.86 |
| *Dipteryx oleifera* | 59.13 | 21.85 | 223086.06 | 2286.05 | 2831.00 | 2978.87 |
| *Dipteryx oleifera* | 104.24 | 36.08 | 154999.35 | 50766.52 | 15825.87 | 15289.08 |
| *Eschweilera antioquensis* | 32.92 | 20.89 | 114447.30 | 2271.38 | 700.24 | 745.29 |
| *Faramea capillipes* | 5.49 | 4.02 | 14191.80 | 0.83 | 5.52 | 3.82 |
| *Garcinia madruno* | 6.10 | 5.25 | 16418.11 | 3.49 | 8.82 | 6.63 |
| *Garcinia madruno* | 9.75 | 12.75 | 4155.88 | 91.88 | 47.15 | 41.22 |
| *Garcinia madruno* | 9.75 | 13.58 | 5835.90 | 70.05 | 44.52 | 43.89 |
| *Guarea sp 2* | 7.32 | 12.16 | 90895.42 | 127.97 | 25.01 | 21.79 |
| *Gustavia dubia* | 19.51 | 22.27 | 125487.31 | 207.31 | 189.36 | 194.10 |
| *Gustavia nana* | 31.09 | 18.50 | 134997.57 | 273.53 | 308.34 | 322.32 |
| *Gustavia superba* | 6.10 | 6.04 | 263.24 | 13.81 | 6.77 | 6.53 |
| *Gustavia superba* | 7.32 | 5.64 | 1309.68 | 17.53 | 8.71 | 15.46 |
| *Gustavia superba* | 9.14 | 6.05 | 1445.13 | 5.88 | 18.43 | 4.86 |
| *Heisteria sp 2* | 12.19 | 11.10 | 1062.49 | 123.55 | 61.68 | 58.79 |
| *Helicostylis tomentosa* | 14.63 | 12.22 | 77585.87 | 49.53 | 71.87 | 69.35 |
| *Helicostylis tomentosa* | 25.60 | 15.54 | 116291.49 | 307.72 | 301.45 | 314.89 |
| *Hernandia didymantha* | 37.19 | 23.40 | 56104.02 | 965.08 | 1081.17 | 605.66 |
| *Hernandia didymantha* | 51.21 | 23.35 | 84925.78 | 1894.59 | 570.61 | 1153.11 |
| *Himatanthus articulatus* | 24.38 | 15.24 | 31635.65 | 420.24 | 240.93 | 249.57 |
| *Himatanthus articulatus* | 34.75 | 25.36 | 33594.89 | 1078.40 | 791.29 | 843.13 |
| *Huberodendron patinoi* | 9.14 | 10.91 | 5464.20 | 30.87 | 26.47 | 23.21 |
| *Hyeronima alchorneoides* | 36.58 | 14.57 | 32972.67 | 638.58 | 430.43 | 454.28 |
| *Hymenaea courbaril* | 60.35 | 26.23 | 9815.15 | 7308.40 | 3247.07 | 3403.25 |
| *Inga pilosula* | 42.67 | 17.14 | 7893.77 | 891.00 | 691.12 | 735.48 |
| *Inga sp 5* | 55.47 | 25.58 | 58923.52 | 4339.67 | 2211.13 | 2340.24 |
| *Iryanthera hostmannii* | 5.06 | 8.77 | 1155.89 | 11.27 | 9.79 | 7.49 |
| *Iryanthera hostmannii* | 21.95 | 13.09 | 17358.51 | 214.03 | 175.45 | 179.17 |
| *Jacaranda copaia* | 42.67 | 25.74 | 4466.18 | 1475.23 | 1020.68 | 1088.62 |
| *Jacaranda copaia* | 60.35 | 24.55 | 8964.85 | 1499.52 | 1957.63 | 2076.65 |
| *Mabea occidentalis* | 4.88 | 8.57 | 175.26 | 10.42 | 8.59 | 6.43 |
| *Magnolia sambuensis* | 36.58 | 17.69 | 103107.03 | 1404.55 | 658.65 | 700.53 |
| *Matisia bracteolosa* | 29.87 | 17.48 | 7918.56 | 620.55 | 359.02 | 377.09 |
| *Myrcia sp 1* | 29.87 | 22.44 | 636.25 | 1625.15 | 557.09 | 591.07 |
| *Naucleopsis glabra* | 7.92 | 9.38 | 2962.86 | 27.87 | 17.93 | 14.99 |
| *Naucleopsis glabra* | 11.58 | 11.09 | 3161.91 | 85.00 | 41.20 | 37.87 |
| *Naucleopsis glabra* | 31.09 | 24.36 | 20738.98 | 4295.57 | 564.58 | 599.15 |
| *Pentaclethra macroloba* | 11.58 | 10.59 | 6664.92 | 17.09 | 46.96 | 43.70 |
| *Perebea xanthochyma* | 17.68 | 15.42 | 11255.92 | 69.70 | 134.72 | 135.63 |
| *Pourouma sp 1* | 7.32 | 6.83 | 18785.72 | 4.03 | 10.18 | 7.83 |
| *Pourouma sp 1* | 18.29 | 17.71 | 92747.17 | 327.31 | 126.53 | 126.93 |
| *Pouteria sp 4* | 11.58 | 8.36 | 34032.82 | 109.84 | 43.59 | 52.66 |
| *Pouteria sp 4* | 13.41 | 10.54 | 15701.38 | 103.19 | 55.73 | 40.28 |
| *Pouteria sp 5* | 11.58 | 10.12 | 29590.96 | 61.73 | 6350.20 | 31.89 |
| *Pouteria sp 5* | 95.71 | 30.11 | 205048.42 | 8889.81 | 35.25 | 6481.83 |
| *Pouteria torta* | 6.10 | 6.78 | 49323.03 | 98.50 | 11.84 | 9.33 |
| *Protium aracouchini* | 6.10 | 11.39 | 5887.11 | 9.18 | 17.34 | 14.43 |
| *Protium aracouchini* | 9.14 | 9.88 | 88217.54 | 75.56 | 26.55 | 23.29 |

| | | | | | | | | |
|---|---|---|---|---|---|---|---|---|
| | | *Zanthoxylum setulosum* | 9.75 | 8.19 | 6700.70 | 16.97 | 28.65 | 25.35 |
| | | *Zanthoxylum setulosum* | 34.14 | 10.97 | 16944.55 | 598.84 | 394.66 | 415.62 |
| | | *Zizyphus strychnifolia* | 4.27 | 6.25 | 64715.04 | 29.04 | 3.66 | 2.32 |
| | | *Aspidosperma sp 6* | 6.71 | 6.47 | 914.23 | 2.32 | 11.92 | 13.34 |
| | | *Pachira quinata* | 15.24 | 7.50 | 33794.56 | 50.09 | 37.58 | 43.59 |
| | | *Bulnesia arborea* | 4.27 | 6.45 | 1998.73 | 15.27 | 5.24 | 5.52 |
| | | *Bulnesia arborea* | 9.75 | 6.51 | 1128.18 | 36.03 | 25.87 | 29.83 |
| | | *Bursera graveolens* | 51.82 | 9.32 | 27573.37 | 767.20 | 599.78 | 634.63 |
| | | *Bursera simaruba* | 12.80 | 7.61 | 59685.33 | 26.44 | 26.33 | 30.37 |
| | | *Caesalpinia glabrata* | 15.85 | 13.99 | 1634.39 | 265.68 | 146.01 | 166.32 |
| | | *Casearia arguta* | 3.05 | 3.42 | 1075.82 | 4.05 | 2.25 | 2.13 |
| | | *Casearia arguta* | 5.49 | 7.33 | 5182.19 | 28.71 | 11.25 | 12.55 |
| | | *Casearia arguta* | 8.53 | 9.52 | 5223.14 | 478.21 | 30.98 | 35.84 |
| | | *Cavanillesia platanifolia* | 185.20 | 19.11 | 166986.82 | 14670.76 | 15381.34 | 11382.47 |
| | | *Cecropia sp 4* | 18.29 | 11.14 | 3169.94 | 28.33 | 78.32 | 90.58 |
| | | *Cecropia sp 4* | 25.97 | 15.82 | 4458.84 | 188.35 | 211.38 | 237.39 |
| | | *Cedrela odorata* | 19.51 | 10.35 | 98133.78 | 216.30 | 97.51 | 112.33 |
| | | *Ceiba pentandra* | 54.25 | 16.33 | 53959.18 | 355.64 | 686.94 | 719.79 |
| | | *Centrolobium paraense* | 24.38 | 14.17 | 31430.37 | 533.22 | 267.05 | 296.65 |
| | | *Centrolobium paraense* | 28.35 | 13.56 | 118385.30 | 1468.82 | 342.54 | 375.53 |
| | | *Cordia sp 1* | 11.58 | 8.76 | 379.03 | 83.33 | 39.90 | 46.29 |
| | | *Cordia sp 1* | 24.38 | 16.45 | 26238.48 | 278.18 | 291.64 | 322.52 |
| | | *Cordia sp 1* | 28.04 | 7.32 | 7615.17 | 35.84 | 175.43 | 198.52 |
| | | *Cordia sp 1* | 28.04 | 12.01 | 5863.90 | 245.05 | 281.93 | 312.33 |
| | | *Guazuma ulmifolia* | 15.24 | 9.30 | 18155.60 | 431.70 | 62.26 | 72.20 |
| | | *Guazuma ulmifolia* | 28.04 | 8.71 | 33170.09 | 368.70 | 183.94 | 207.77 |
| | | *Hura crepitans* | 75.59 | 15.07 | 63691.70 | 167.66 | 1759.04 | 1700.92 |
| | | *Inga sp 10* | 15.24 | 10.66 | 70254.42 | 104.82 | 87.17 | 100.64 |
| **Acalypha sp. and Guazuma ulmifolia** | **Bttf3/Asp-Gul** | *Lecythis minor* | 9.14 | 6.43 | 4693.51 | 1.78 | 10.98 | 12.24 |
| | | *Lecythis minor* | 67.67 | 20.88 | 125914.91 | 2175.82 | 1427.85 | 1408.14 |
| | | *Maclura tinctoria* | 54.86 | 16.45 | 8995.05 | 1779.63 | 1524.73 | 1494.56 |
| | | *Ochroma pyramidale* | 51.21 | 15.30 | 39991.15 | 873.52 | 482.59 | 518.20 |
| | | *Pouteria sp 8* | 39.62 | 12.05 | 71437.87 | 681.19 | 455.68 | 491.11 |
| | | *Pouteria sp 9* | 14.02 | 11.14 | 105958.29 | 75.78 | 68.38 | 79.23 |
| | | *Pseudobombax septenatum* | 12.19 | 9.48 | 3374.69 | 10.00 | 22.71 | 26.11 |
| | | *Pseudobombax septenatum* | 42.06 | 10.36 | 27447.70 | 74.01 | 307.41 | 339.02 |
| | | *Pterocarpus acapulcensis* | 85.95 | 13.62 | 94264.93 | 4510.72 | 2424.16 | 2269.39 |
| | | *Samanea saman* | 80.47 | 17.62 | 267915.25 | 8289.91 | 2374.36 | 2227.61 |
| | | *Samanea saman* | 88.39 | 14.96 | 206779.90 | 4504.60 | 2433.84 | 2277.51 |
| | | *Sapium glandulosum* | 29.26 | 12.11 | 52370.74 | 456.37 | 229.76 | 257.07 |
| | | *Spondias mombin* | 8.53 | 7.03 | 25688.21 | 9.10 | 13.56 | 15.28 |
| | | *Spondias mombin* | 63.40 | 17.24 | 226782.11 | 1131.54 | 1386.05 | 1370.64 |
| | | *Sterculia apetala* | 95.71 | 20.49 | 727187.96 | 2126.99 | 3708.25 | 3312.87 |
| | | *Swartzia simplex* | 5.49 | 5.56 | 2951.80 | 4.58 | 9.80 | 10.84 |
| | | *Tabebuia rosea* | 32.31 | 9.23 | 188089.14 | 223.29 | 216.94 | 243.35 |
| | | *Tabebuia sp 1* | 31.70 | 10.83 | 163487.46 | 155.98 | 203.48 | 228.88 |
| | | *Tabernaemontana cymosa* | 16.46 | 8.33 | 9638.95 | 70.60 | 58.63 | 68.02 |
| | | *Tabernaemontana cymosa* | 18.29 | 8.52 | 7586.28 | 167.28 | 72.79 | 84.28 |
| | | *Tectona grandis* | 15.24 | 14.09 | 41158.49 | 250.57 | 39.98 | 46.39 |
| | | *Tectona grandis* | 21.34 | 16.34 | 73923.97 | 727.00 | 85.23 | 98.44 |
| | | *Terminalia oblonga* | 38.40 | 21.34 | 155083.75 | 1898.86 | 1014.02 | 1030.48 |
| | | *Terminalia oblonga* | 45.11 | 19.19 | 153736.66 | 2308.70 | 1258.80 | 1255.63 |
| | | *Terminalia oblonga* | 51.82 | 21.97 | 91797.37 | 4298.41 | 1909.44 | 1831.56 |
| | | *Triplaris americana* | 16.46 | 8.26 | 18354.83 | 29.15 | 83.22 | 96.16 |
| | | *Xylosma sp 2* | 17.68 | 10.42 | 11366.03 | 219.07 | 103.75 | 119.34 |
| | | *Zanthoxylum setulosum* | 23.16 | 10.96 | 7058.47 | 291.92 | 186.73 | 210.79 |

**Table 2 | Comparison of biomass estimates for the X matrix (DBH; height; biomass of the stem, main branches, and secondary branches or twigs; leaf biomass; and total biomass, using three-dimensional modeling) and the Y matrix (same structural variables as X, plus the biomass of organs estimated with allometric equations) (*13*).**

| Forests in superhumid to humid climates | | | | | |
|---|---|---|---|---|---|
| **DBH range** | **Size nx and ny** | **$H_0$** | **- 2log (λR)** | **[x1- α, + ∞)** | **p** |
| 0–20 cm | 81 | | 2511.54 | | |
| 20–40 cm | 44 | | 1861.10 | | |
| 40–60 cm | 15 | rejected | 646.92 | [24.99, + ∞) | 0.00 |
| 60–100 | 14 | | 603.82 | | |
| Higher than 100 cm | 8 | | 335.42 | | |
| **Forests in semihumid climates** | | | | | |
| 0–20 cm | 24 | | 705.12 | | |
| 20–40 cm | 14 | rejected | 470.75 | [24, 99, + ∞) | 0.00 |
| 40–60 cm | 7 | | 238.89 | | |
| Higher than 60 cm | 8 | | 413.65 | | |

**Table 3 | Distribution of carbon stocks as storage intervals per hectare, and vegetation formations in pantropical regions.**

| Class | Carbon stocks range (Mg C $ha^{-1}$) | Type of vegetation formation | Carbon stocks (Pg) | Flat surface (ha) | Relationship of carbon stocks and flat surface ($ha^{-1}$) | Percentage |
|---|---|---|---|---|---|---|
| 5 | 478–605 | **Highly conserved forests with multiple layers** | 285.03 | 531,449,420.52 | 531.05 | 39.37 |
| 4 | 336–478 | | 281.76 | 668,896,502.68 | 417.09 | 38.91 |
| 2 | 58–183 | **Very disturbed forest, scrub forest, and tall scrub** | 105.54 | 419,412,183.77 | 249.16 | 14.57 |
| 3 | 183–336 | **Secondary forests, moderately disturbed forest and woodland, interspersed with scrubs** | 40.66 | 350,688,503.17 | 114.80 | 5.61 |
| 1 | 1–58 | **Savannah vegetation dominant, forest fragments, or completely degraded forests, low shrubs** | 10.96 | 828,606,996.75 | 13.10 | 1.51 |
| **Total** | | | 723.97 | **2,799,053,606.91** | | |

Mg: megagrams; C: carbon; $ha^{-1}$: per hectare; Pg: petagrams.

## Table 4 | Countries with high carbon stocks, estimated using the methodology proposed in this study.

| Countries by amount of carbon stocks | | | | Countries by carbon stocks per unit area | | | |
|---|---|---|---|---|---|---|---|
| **Country** | **Country area ($km^2$)** | **Carbon stock (Pg)** | **Carbon per $km^2$** | **Country** | **Country area ($km^2$)** | **Carbon stock (Pg)** | **Carbon per $km^2$** |
| **Brazil** | 8,507,128 | 194.24 | 22,833.45 | **Equatorial Guinea** | 27,085.30 | 1.16 | 43,107.19 |
| **Zaire** | 2,337,027 | 71.95 | 30,787.99 | **Brunei** | 5769.53 | 0.24 | 42,034.99 |
| **Indonesia** | 1,910,842 | 64.41 | 33,710.66 | **Papua New Guinea** | 466,161.18 | 19.45 | 41,741.36 |
| **Colombia** | 1,141,962 | 37.34 | 32,701.25 | **French Guiana** | 83,811.13 | 3.46 | 41,370.25 |
| **Peru** | 1,296,912 | 36.12 | 27,854.89 | **Guyana** | 211,241.29 | 8.60 | 40,741.44 |
| **Bolivia** | 1,090,353 | 28.82 | 26,439.54 | **Suriname** | 145,497.5 | 5.91 | 40,656.40 |
| **Venezuela** | 916,561 | 23.63 | 25,782.34 | **Gabon** | 261,688.70 | 10.53 | 40,249.97 |
| **Papua New Guinea** | 466,161 | 19.45 | 41,741.36 | **Laos** | 230,566.09 | 9.23 | 40,050.80 |
| **Mexico** | 1962939 | 15.29 | 7792.11 | **Liberia** | 96296.03 | 3.63 | 37735.46 |
| **Myanmar (Burma)** | 669,820.87 | 12.96 | 19,354.38 | **Malaysia** | 330,269.59 | 12.22 | 37,009.20 |

$Km^2$: square kilometers; Pg: petagrams.

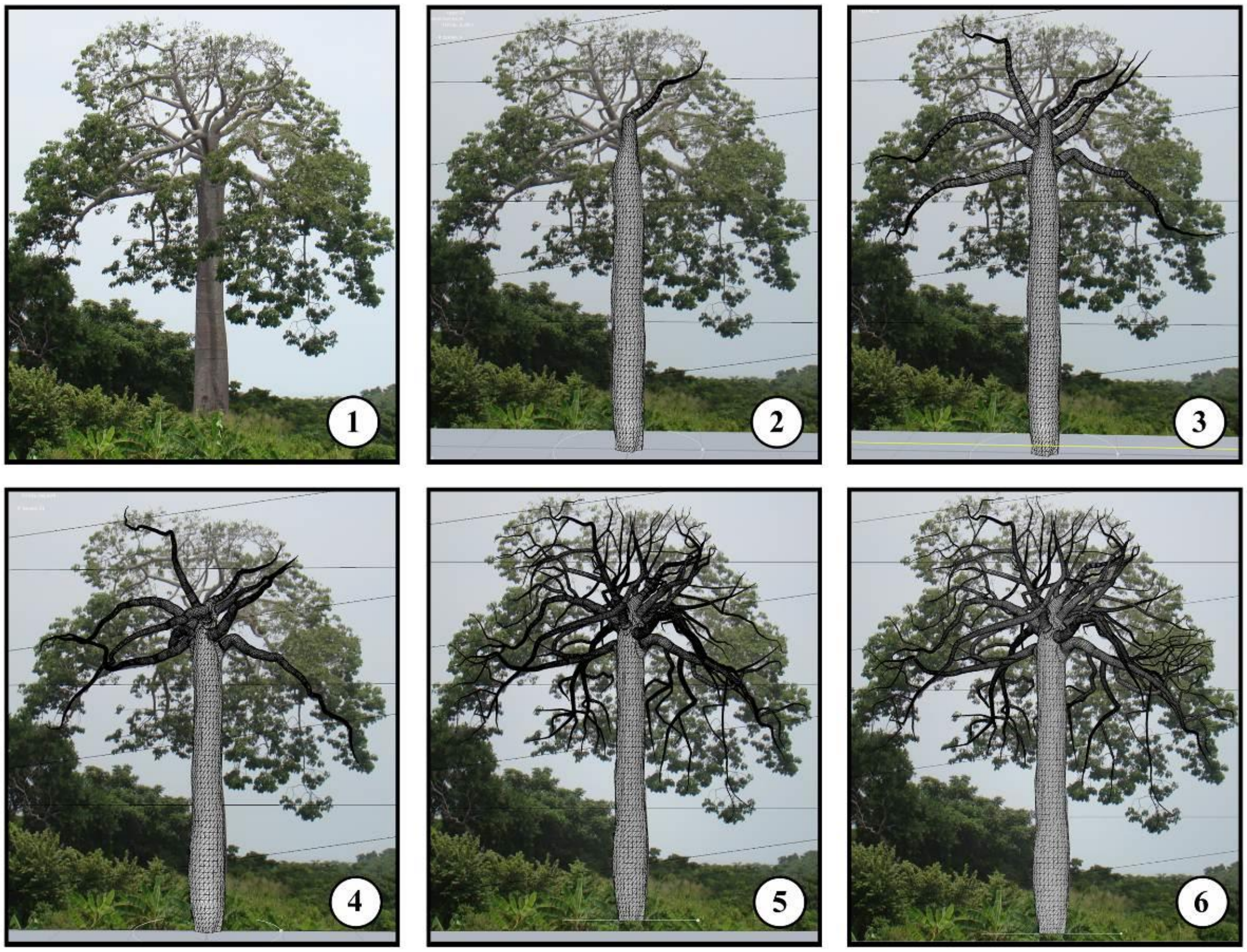

**Figure 1.** Building process of bounded three-dimensional model of the fourth-level branching system in an individual of *Cavallinesia platanifolia* (Güipo, Volado, Macondo). The fifth and sixth levels were modeled separately to accelerate the calculation process. (1), the individual to be modeled; (2) and (3), the first surfaces without boundaries (perspective view); and (4), (5), and (6), the stem and branch surfaces with size and position corrections (bounded model, XZ side view). Besides the traditional coverage measures, we used a lateral photo (YZ) or a side-angle shot to correct the nodes position of the Bezier curves and obtain optimal results. For preparation of the figure, SpeedTree 6.1.1 was used.

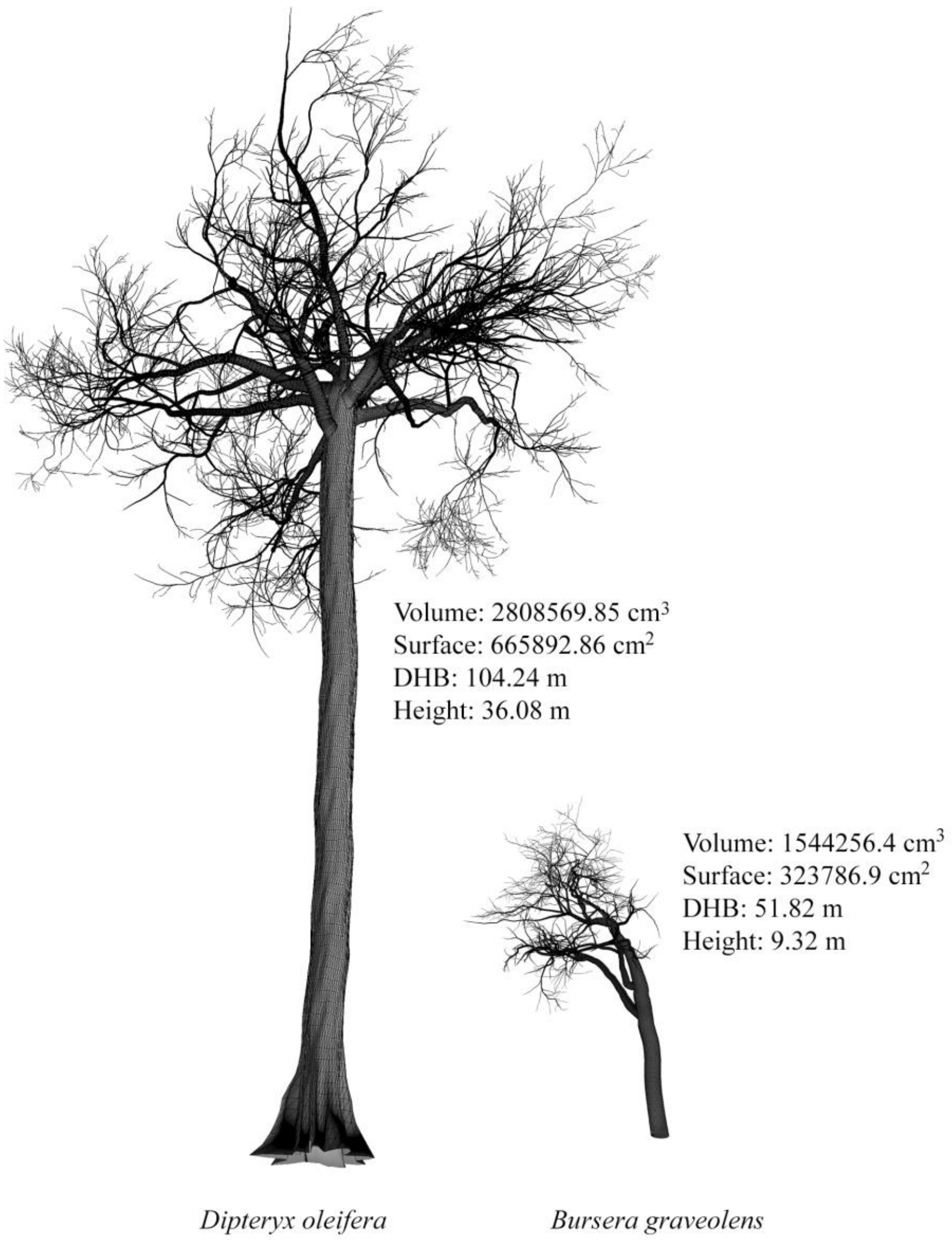

**Figure 2.** Mesh models of individuals of *Dipteryx oleifera* with DBH of 104.24 cm and height of 36.08 m and *Bursera graveolens* with DBH of 51.82 cm and height of 9.32 m. In general, branches of large trees below two-thirds of its height indicate anthropogenic disturbance because they result from the removal by humans of competitors located in lower strata. Table shows the results of the numerical integration method by Mirtich (1996). For code implementation and preparation of the figure, MATLAB 2011b was used.

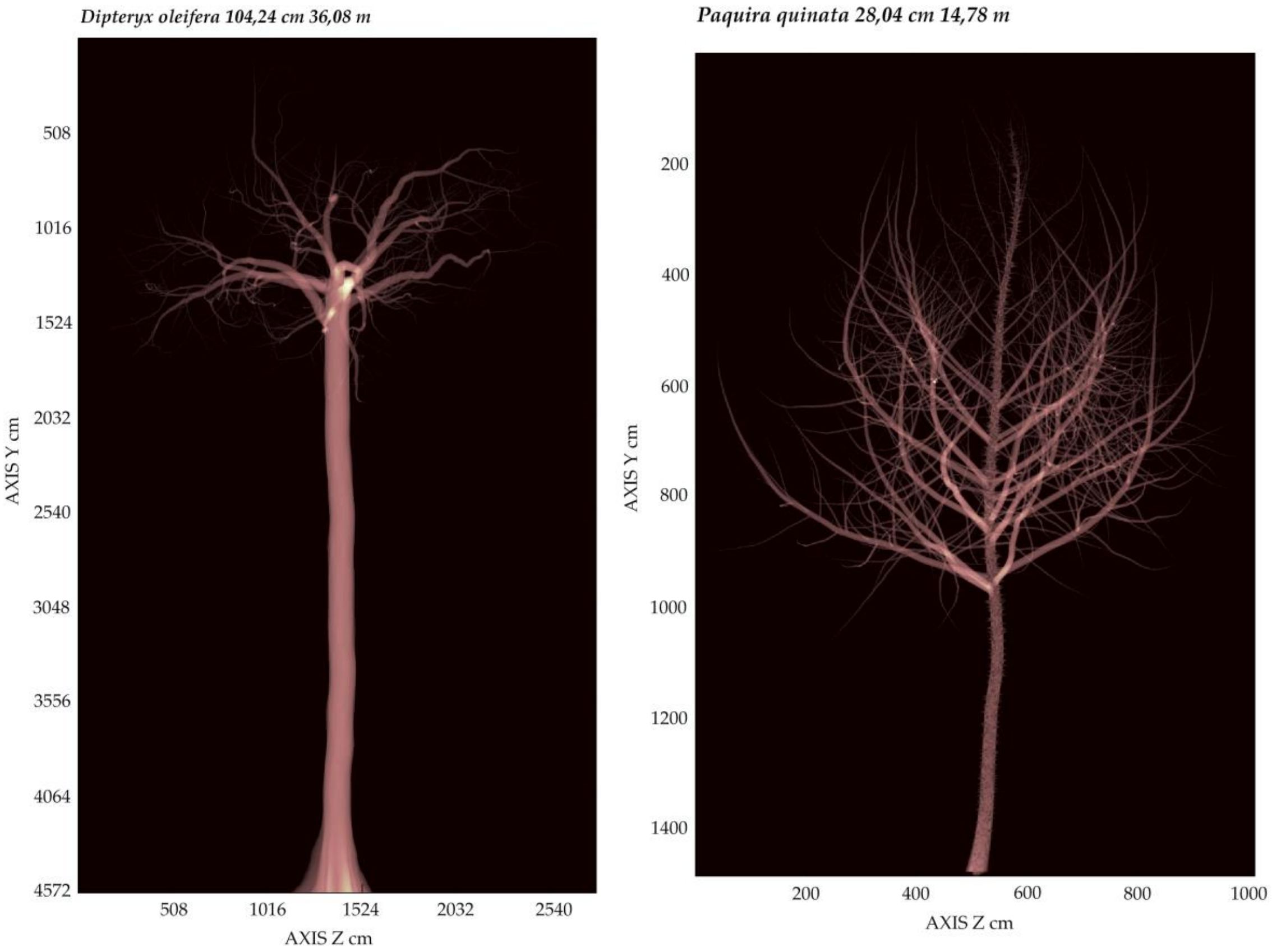

**Figure 3**. Voxel models of individuals of *Dipteryx oleifera* with DBH of 104.24 cm and height of 36.08 m, and *Paquira quinata* with DBH of 28.04 cm and height of 14.78 m, generated by modifying Aitkenhead's (2011) code. MATLAB 2011b was used for the code implementation and preparation of the figure.

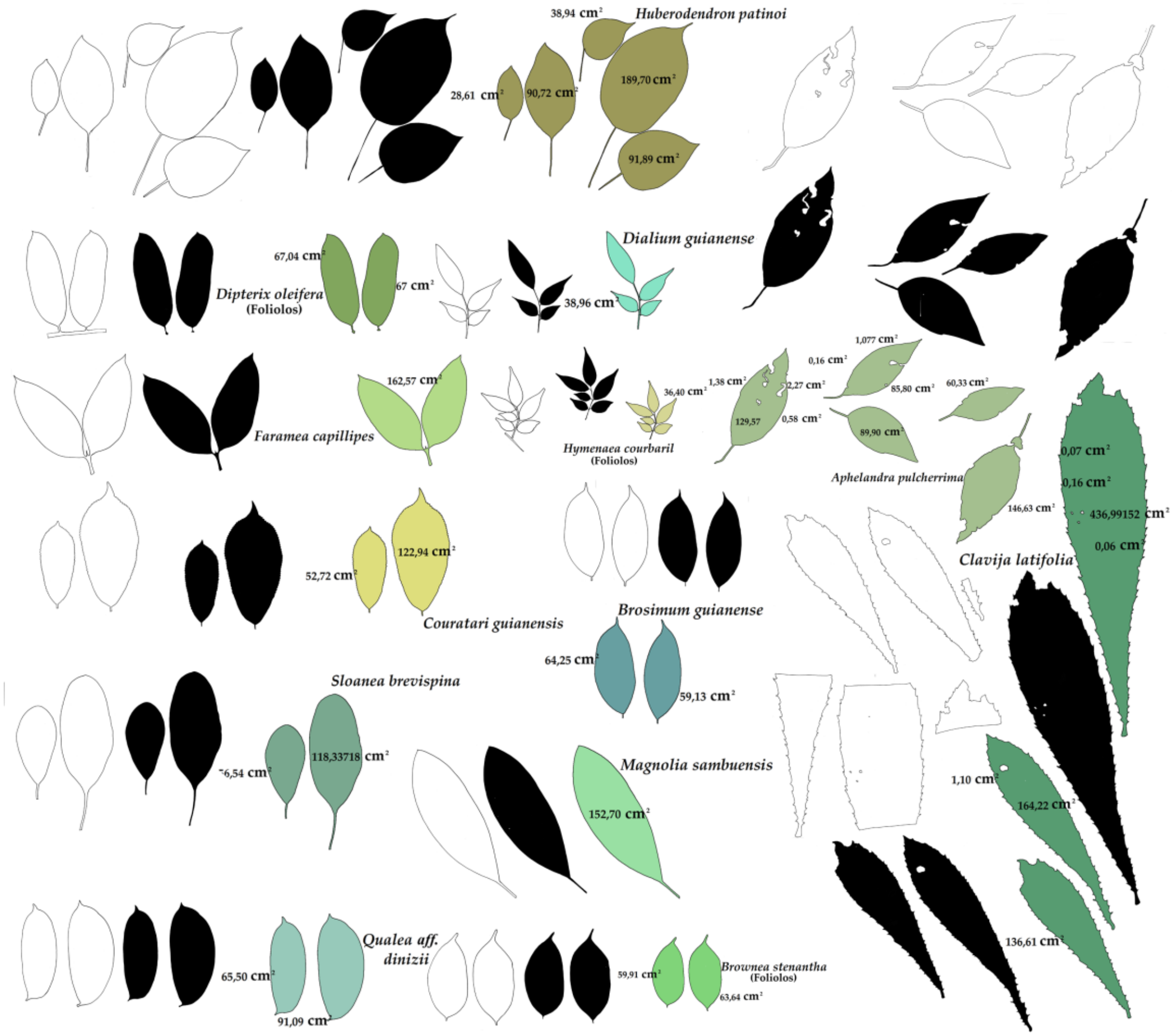

**Figure 4**. Ink silhouette, color filling and results of vectorization of leaf blades, some support structures and some layer reconstruction. Some of the important leaf layer features are seen, including herbivory. Information on complex or divided leaves was carefully reconstructed to facilitate their illustration in the final matrices. Black color filling is the aspect that information acquires in raster format (.TIF) of one bit per pixel. The attribute table shows which object belongs to a higher order structure (composite leaf), and which structure belongs to a layer segment. Results were rendered using GIMP 2.6 and GRASS 6.4.

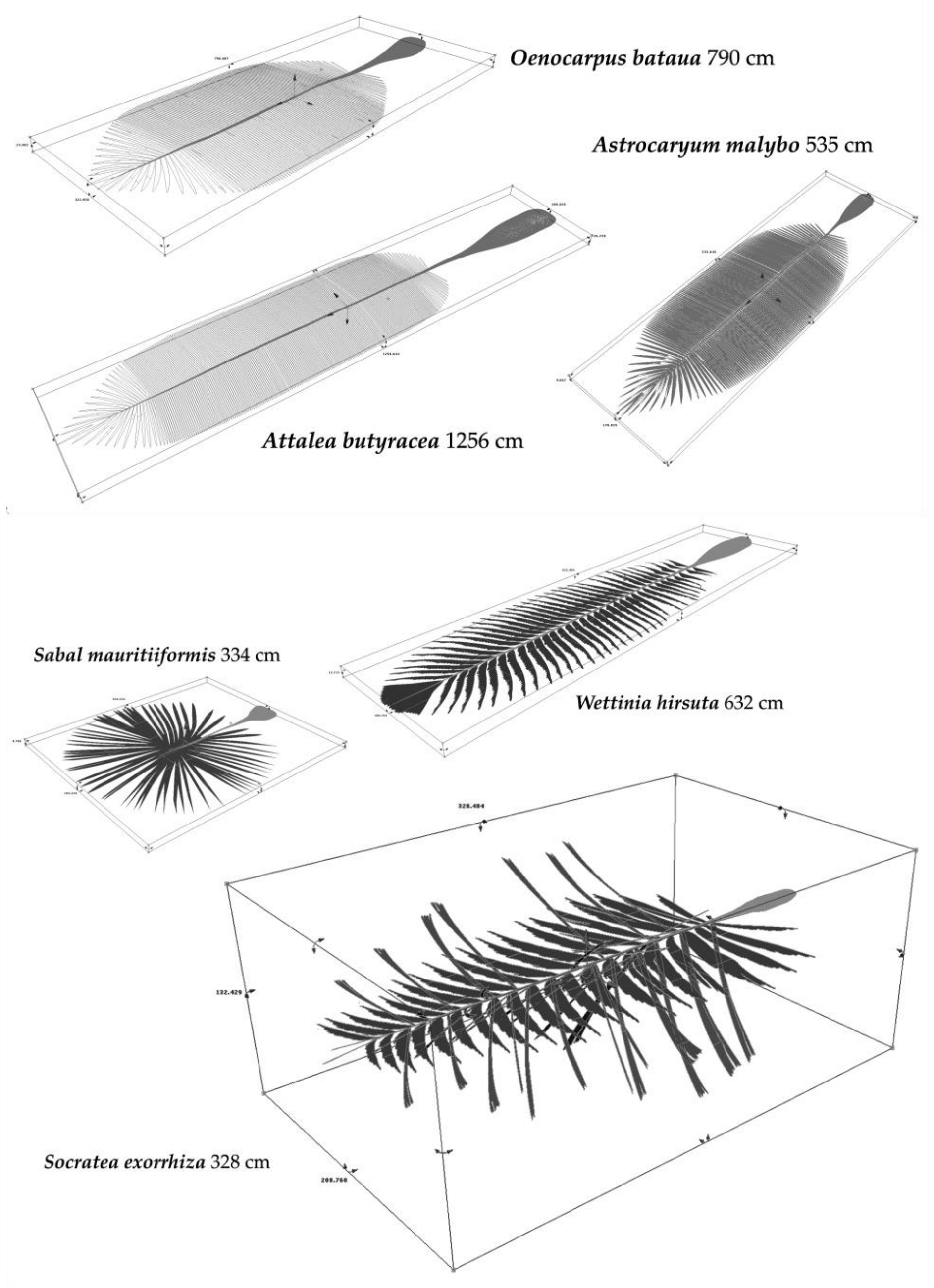

**Figure 5**. Results of palm species modeling using Maya 2009 software (Autodesk, 2008). Model measurements in centimeters are shown.

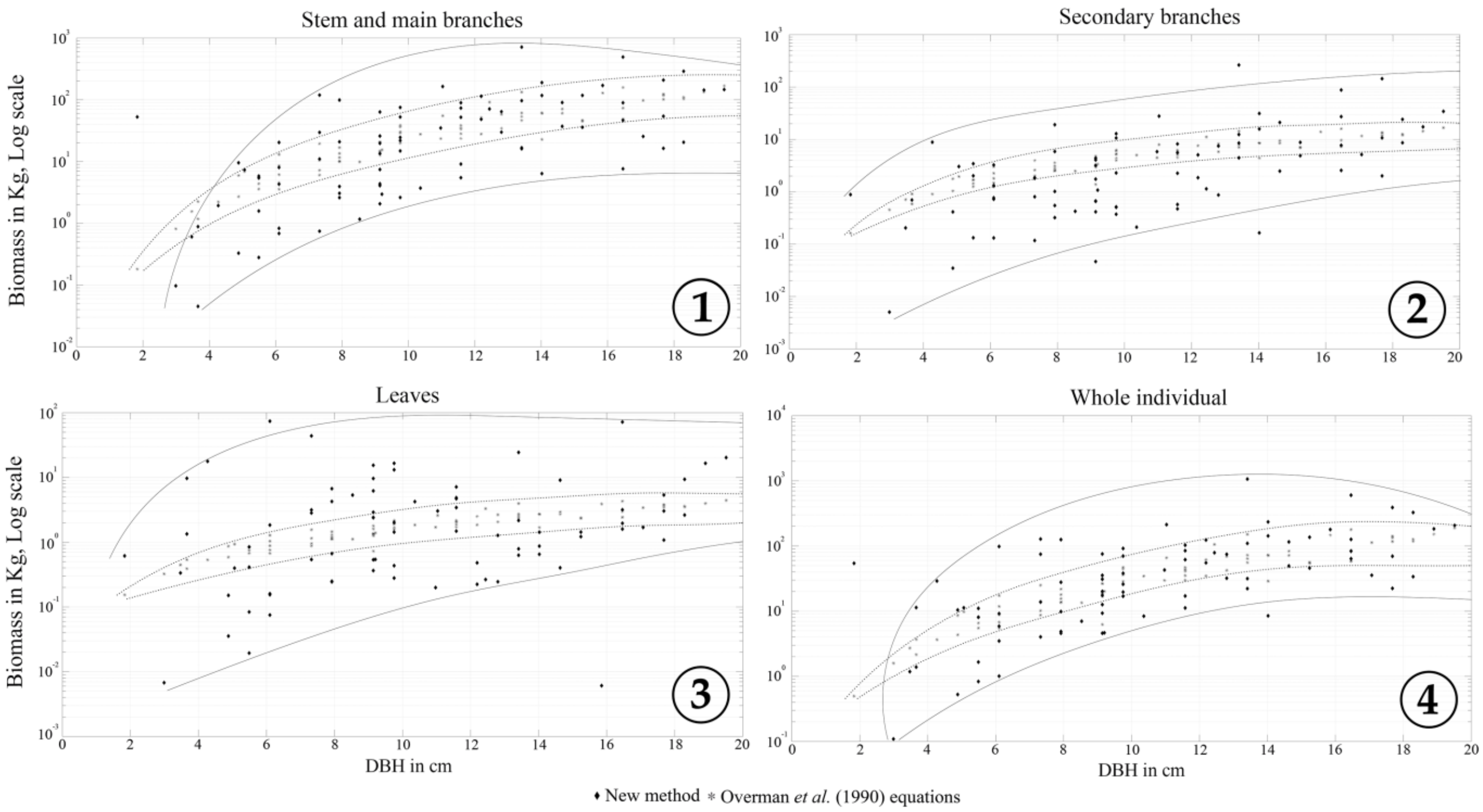

**Figure 6.** Comparison of the estimated biomass for established forests in superhumid to humid climates studied using three-dimensional modeling of individuals (dark rhomboid) and the traditional allometric method published by Overman *et al.* (1990) (grey dots) for the various organs studied and for individuals with DBH between 1 and 20 cm. For preparation of the figure, the software MATLAB 2011b was used.

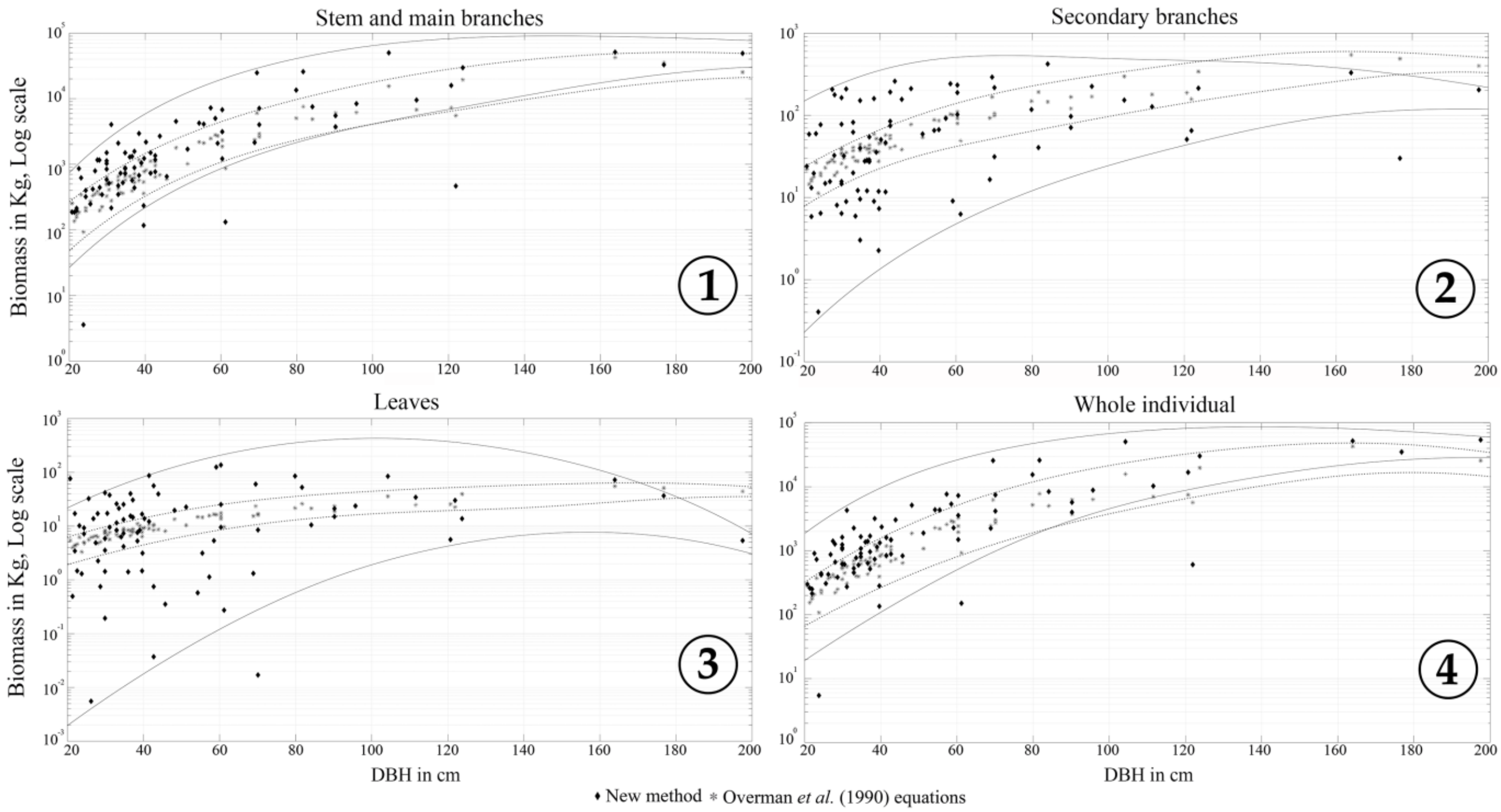

**Figure 7.** Comparison of the estimated biomass for established forests in superhumid to humid climates studied using three-dimensional modeling of individuals (dark rhomboid) and the traditional allometric methodology published by Overman *et al.* (1990) (grey dots) for various organs studied with DBH range 20 and 200 cm. The software MATLAB 2011b was used for preparation of the figure.

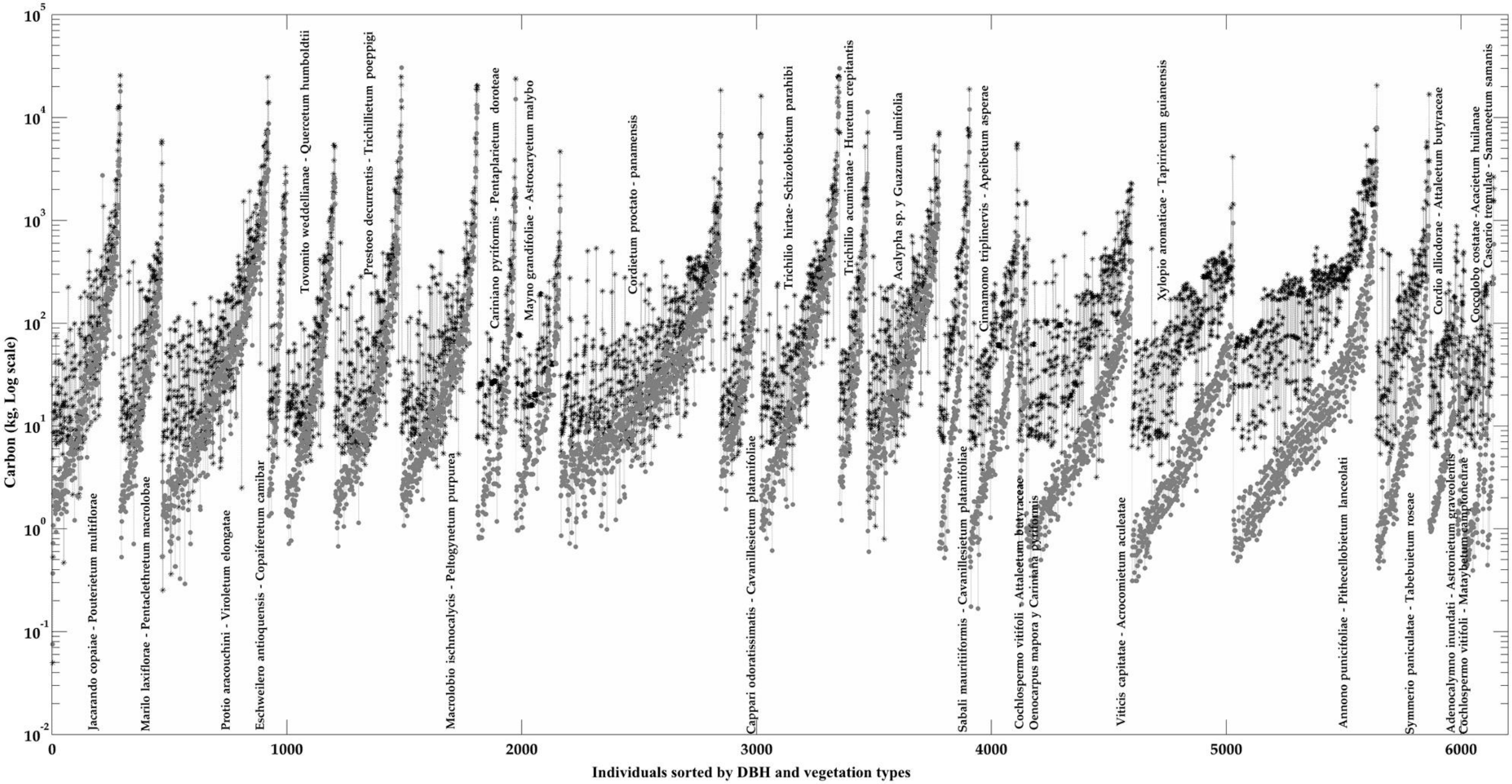

**Figure 8.** Results in kilograms (logarithmic scale) of the estimated carbon content for more than 6,000 individuals of different species belonging to 27 vegetation types in the studied area. Results obtained using three-dimensional modeling (dark asterisk) and calibration using artificial neural networks are contrasted, and the values were found by applying equations of Chave. *et al.* (2005)(grey dots). MATLAB 2011b software was used to prepare the figure.

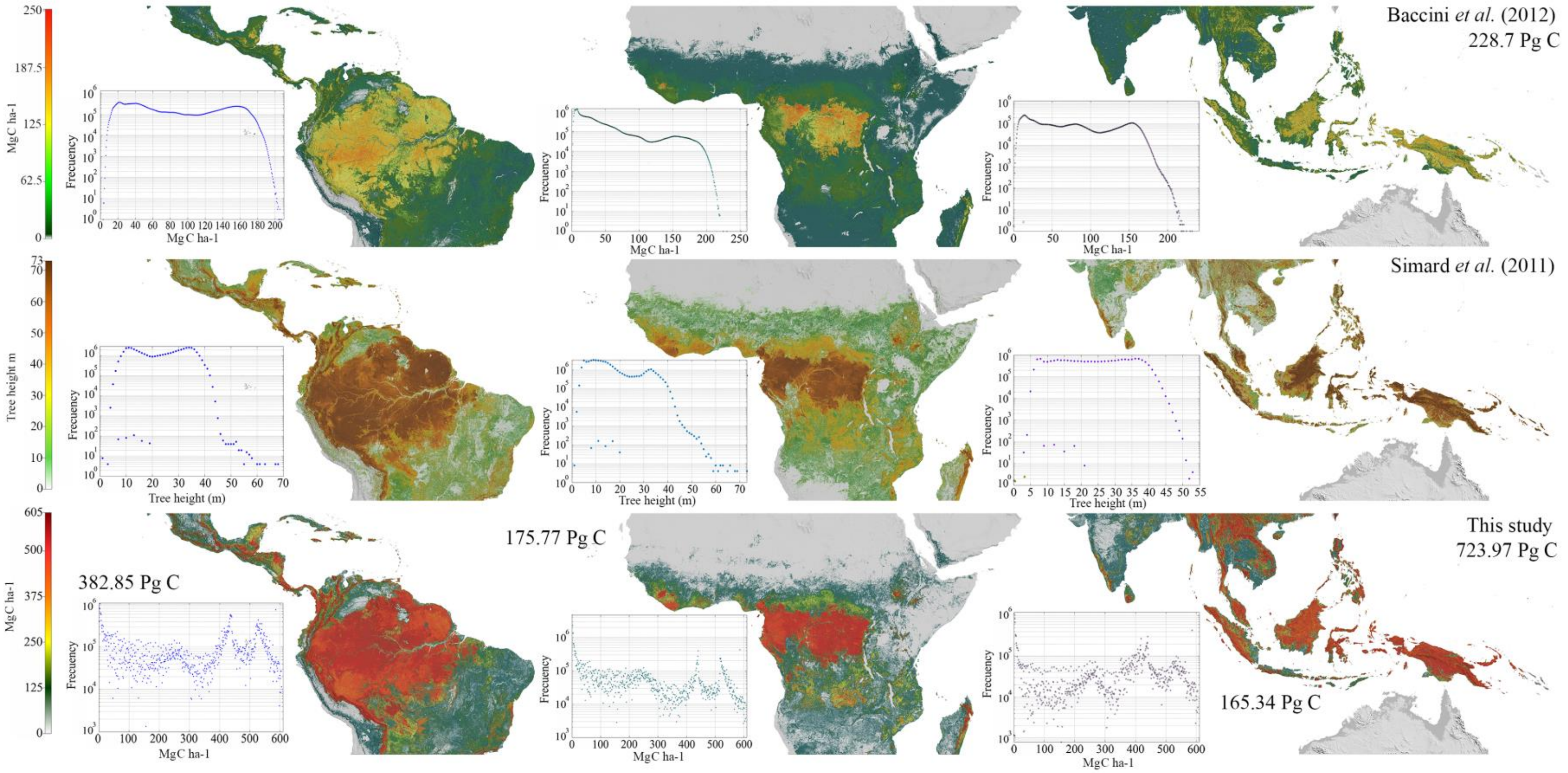

**Figure 9.** Pantropical carbon maps. Top row: Values are in Mg C ha$^{-1}$ per continent as in Baccini *et al.* (2012). Map data were produced by Baccini *et al.* (2012). Middle row: Distribution of tree height by continent, provided by Simard *et al.* (2011). Map data were produced by Simard *et al.* (2011). Bottom row: Results of calibration of the pantropical carbon map based on our methodology. Note the dispersal of data attributable to incorporating different architectures in estimating carbon per hectare. Map is the result of modeling using neural network back-propagation for results obtained in this study and information from Baccini *et al.* (2012) (top) and Simard et al. (2011) (middle). MATLAB 2011b was used by H. Arellano to process and present the distribution of frequencies in Mg C ha$^{-1}$ (logarithmic scale). The results were rendered by H. Arellano in GRASS 6.4, and the composition was made with GIMP 2.6.

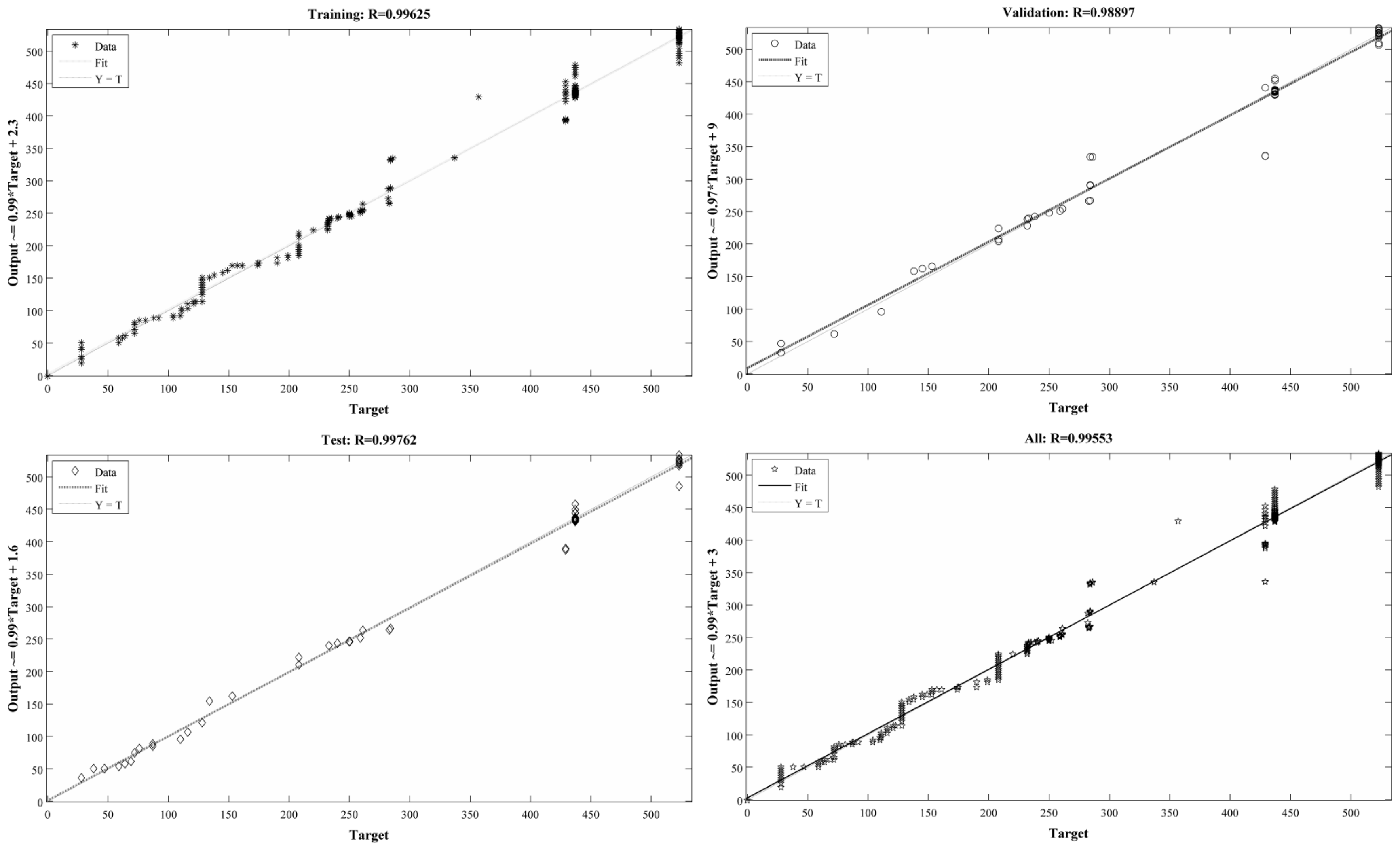

**Figure 10.** Data adjustment results from an artificial neural network. Data were transformed by a logarithmic sigmoidal function with linear output. In addition to the adjustments made to the data used in training, the system randomly chose a group of records to test the validation. Another similar test was performed using random values located in the range of DBH evaluated. The end result is the correlation coefficient among all tests. This approach allowed the selection of the neural models that gave the best explanation of evaluated records. MATLAB 2011b software was used to prepare the figure.

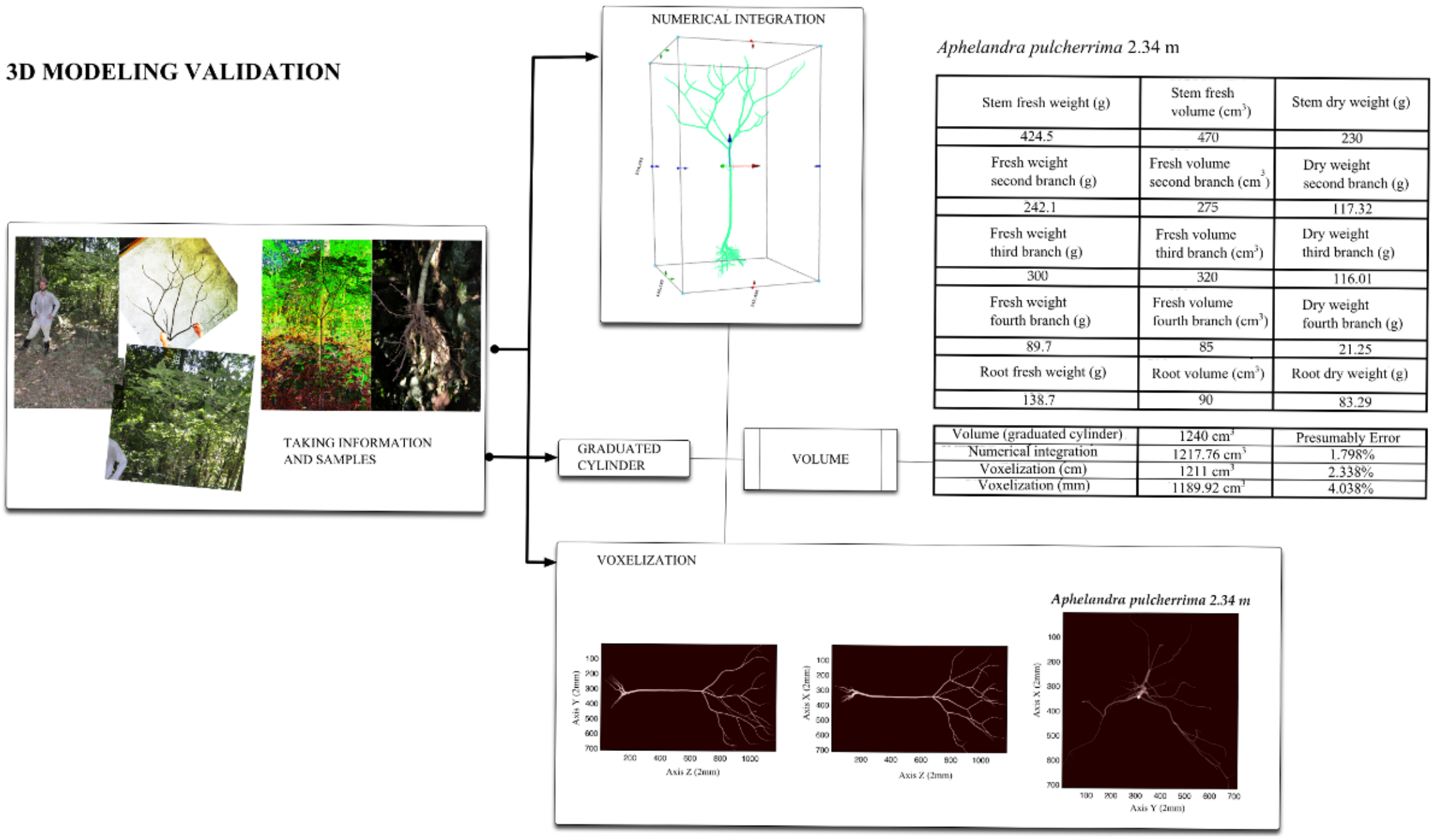

**Figure 11**. Voxelization method validation process. For preparation of the figure, the software MATLAB 2011b and GIMP 2.6 were used.

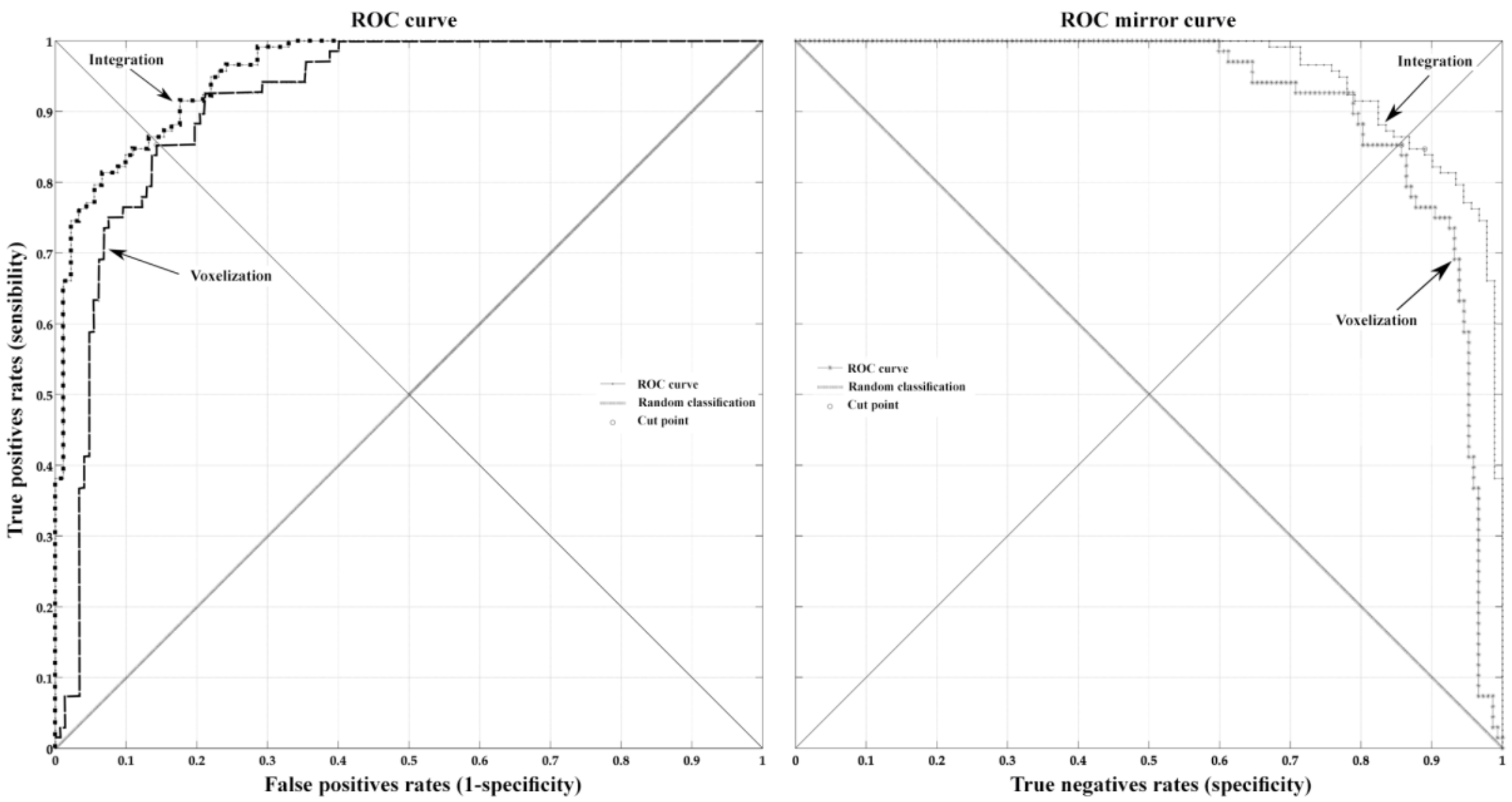

**Figure 12**. Rates of false positives and true negatives regarding the rates for the resultant true positives for the voxelization and numerical integration methodologies. For preparation of the figure, the software MATLAB 2011b was used.